\documentclass[useAMS,usenatbib]{mn2e}
\usepackage{titlesec, color}
\usepackage{graphicx}
\usepackage{float}
\usepackage{aashack}
\usepackage{amssymb}
\usepackage{booktabs}
\usepackage{caption}
\usepackage{natbib}
\usepackage[pdftex,
            colorlinks = true,
            linkcolor = blue,
            urlcolor  = blue,
            citecolor = blue,
            linktocpage = true,
            anchorcolor = blue]{hyperref}

\title[MATLAS Dwarfs]{Newly discovered dwarf galaxies in the MATLAS low density fields}
\author[R Habas et al.]{Rebecca Habas$^{1,2}$\thanks{E-mail:
rebecca.habas@gmail.com}, Francine R.\ Marleau$^{1}$, Pierre-Alain Duc$^{2}$, Patrick R.\ Durrell$^{3}$, \newauthor Sanjaya Paudel$^{4}$, M{\'e}lina Poulain$^{1}$, Rub{\'e}n S{\'a}nchez-Janssen$^{5}$, Sreevarsha Sreejith$^{6}$,\newauthor Joanna Ramasawmy$^{7}$, Bryson Stemock$^{3}$, {Christopher Leach}$^{3}$, \newauthor Jean-Charles Cuillandre$^{8}$, {Stephen Gwyn}$^{9}$, {Adriano Agnello}$^{10}$, Michal B{\'ilek}$^{2}$, \newauthor 
J\'er\'emy Fensch$^{11,12}$, Oliver M{\"u}ller$^{2}$, Eric W. Peng$^{13}$, Remco F.J. van der Burg$^{11}$\\
$^{1}$Institut f{\"u}r Astro- und Teilchenphysik, Universit{\"a}t Innsbruck, Technikerstra{\ss}e 25/8, Innsbruck, A-6020, Austria \\
$^{2}$Observatoire Astronomique, Universit{\'e} de Strasbourg, CNRS, 11, rue de l'Universit{\'e}. F-67000 Strasbourg, France \\
$^{3}$Department of Physics and Astronomy, Youngstown State University, Youngstown, OH, 44555, USA\\
$^{4}$Department of Astronomy and Center for Galaxy Evolution Research, Yonsei University, Seoul 03722\\
$^{5}$UK Astronomy Technology Centre, Royal Observatory Edinburgh, Blackford Hill, Edinburgh EH9 3HJ, UK\\
$^{6}$Universite Clermont Auvergne, CNRS/IN2P3, LPC, F-63000 Clermont-Ferrand, France\\
$^{7}$Centre for Astrophysics Research, School of Physics, Astronomy and Mathematics, University of Hertfordshire, College Lane,\\ Hatfield AL10 9AB, UK \\ 
$^{8}$IRFU, CEA, Universit\'e Paris-Saclay, Universit\'e Paris Diderot, AIM, Sorbonne Paris Cit\'e, CEA, CNRS, Observatoire de Paris, \\ PSL Research University, F-91191 Gif-sur-Yvette Cedex, France \\
$^{9}$NRC Herzberg Astronomy and Astrophysics, 5071 West Saanich Road, Victoria, BC, V9E 2E7, Canada \\
$^{10}$DARK, Niels Bohr Institute, University of Copenhagen, Lyngbyvej 2, 2100 Copenhagen, Denmark \\
$^{11}$European Southern Observatory, Karl-Schwarzschild-Str. 2, D-85748 Garching, Germany\\
$^{12}$Univ. Lyon, ENS de Lyon, Univ. Lyon 1, CNRS, Centre de Recherche Astrophysique de Lyon, UMR5574, F-69007 Lyon, France \\
$^{13}$Department of Astronomy and Kavli Institute for Astronomy and Astrophysics, Peking University, Beijing, 100871, China \\
}
\begin{document}

\date{}

\pagerange{\pageref{firstpage}--\pageref{lastpage}} \pubyear{2018}

\maketitle

\label{firstpage}

\begin{abstract}
We present the photometric properties of 2210 newly identified dwarf galaxy candidates in the MATLAS fields. The Mass Assembly of early Type gaLAxies with their fine Structures (MATLAS) deep imaging survey mapped $\sim$142~deg$^2$ of the sky around nearby isolated early type galaxies using MegaCam on the Canada-France-Hawaii Telescope, reaching surface brightnesses of $\sim$ 28.5~--~29 in the \textit{g}-band. The dwarf candidates were identified through a direct visual inspection of the images and by visually cleaning a sample selected using a partially automated approach, and were morphologically classified at the time of identification. Approximately 75\% of our candidates are dEs, indicating that a large number of early type dwarfs also populate low density environments, and 23.2\% are nucleated. Distances were determined for 13.5\% of our sample using pre-existing $z_{spec}$ measurements and HI detections. We confirm the dwarf nature for 99\% of this sub-sample based on a magnitude cut $M_g = -18$.  Additionally, most of these ($\sim90$\%) have relative velocities suggesting that they form a satellite population around nearby massive galaxies rather than an isolated field sample. Assuming that the candidates over the whole survey are satellites of the nearby galaxies, we demonstrate that the MATLAS dwarfs follow the same scaling relations as dwarfs in the Local Group as well as the Virgo and Fornax clusters. We also find that the nucleated fraction increases with $M_g$, and find evidence of a morphology-density relation for dwarfs around isolated massive galaxies.

\end{abstract}
\begin{keywords}
galaxies: dwarf --- galaxies: photometry --- galaxies: structure --- galaxies, fundamental parameters.
\end{keywords}

\section{Introduction}
The properties of low surface brightness dwarf galaxies in low-density environments are poorly constrained observationally. Deep observations over large portions of the sky are required to detect a large sample of such galaxies, which is expensive in terms of telescope time. As a result, most studies focus on dwarfs in the Local Group, where they can be studied in detail, selected nearby groups, and those in nearby clusters (in particular Virgo, Coma, and Fornax) where dwarfs can be found in greater numbers and their association with the cluster can be used to constrain their distances. By comparison, relatively few observing programs have focused on dwarf populations around otherwise isolated massive galaxies.

However, dwarfs in low density environments provide a crucial sample to test whether or not the dwarfs in the Local Group are representative of dwarfs in general. In terms of their star formation histories, \citet{Weisz11} have confirmed that dwarfs in the Local Group are indeed similar to those found at $d\lesssim4$~Mpc in the ACS Nearby Galaxy Survey Treasury Program (ANGST; \citealt{Dalcanton09}). However, it remains an open question if other properties of Local Group dwarfs, such as the number and distribution of dwarf satellites, are also representative. For example, it is known that bright satellites like the Large Magellanic Cloud (LMC) are relatively rare; \citet{Tollerud11} found that only $\sim$12\% of isolated Milky Way analogues in the Sloan Digital Sky Survey (SDSS; \citealt{York00}) have an LMC-like satellite within a projected distance of 75~kpc. More generally, any variation in the number of dwarf satellites around massive galaxies is not well constrained observationally, and this may have important implications for the missing satellite problem \citep{Kauffmann93,Klypin99,Moore99} and/or constraining cosmological simulations. Observations have revealed, however, that dwarfs around the Milky Way and Andromeda lie in a thin planes around their respective hosts (e.g., \citealt{Pawlowski12,Ibata13,Santos-Santos19}); several other satellite planes have since been identified around M81, NGC~3109, Centaurus~A, M101, and potentially M83 \citep{Chiboucas13,Bellazzini13,Tully15,Muller17,Muller18a,Muller18b}, suggesting that the Local Group satellites are typical, but this remains a challenge to explain in the standard cosmological model (see \citealt{Pawlowski18} and \citealt{Bullock17} for reviews, and \citealt{Kroupa15} or \citealt{Bilek18} for an alternative explanation).

Dwarfs in low density environments are also important to understand evolutionary trends in low mass galaxies within clusters. There is growing evidence that dwarfs are pre-processed in group environments --- impacting both their star formation rates and morphologies --- before they fall into a cluster potential (e.g., \citealt{Cortese06,Cybulski14,Roberts17,Mendelin17}).

Over the years there has been a growing interest in observing dwarfs within low density environments. Some groups have focused on building catalogues of nearby galaxies  --- including dwarfs --- with a range of consistent observables \citep{Impey96,Schombert01, Hunter04, Karachentsev04}. Others have taken advantage of large, public data sets to search for new dwarfs. For example, \citet{Roberts04a} investigated number counts of dwarf galaxies in the Millennium Galaxy strip to probe variations in the dwarf-to-giant ratio in different environments. They found a dwarf-to-giant ratio of 6:1, significantly smaller than the 20:1 ratio in the Virgo cluster. \citet{Geha12} used data from the SDSS to study star formation rates of almost 3000 dwarf galaxies, concluding that essentially all isolated dwarfs (those without a neighbouring galaxy within 1~Mpc) have ongoing star formation. More recently, \citet{Ann17} supplemented SDSS DR7 spectroscopic distances with redshifts from other sources to confirm $\sim$2600 dwarf galaxies in the Catalogue of Visually Classified Galaxies (CVCG; \citealt{Ann15}) in the nearby Universe; the catalogue also includes morphological classifications, and the morphologies of the dwarf satellites appear to correlate with the morphology of the host galaxy, such that more dEs are found around massive early type galaxies.

Several deep optical imaging observing programs have begun collecting data in recent years, either for the express purpose of detecting new dwarf galaxies or with this as a secondary science goal. These include large sky surveys such as the Panoramic Survey Telescope and RApid Response System (PAN-STARRS1; \citealt{Chambers16}), the Dark Energy Survey (DES; \citealt{DES05}), Survey of the MAgellanic Stellar History (SMASH; \citealt{Nidever17}), and the VLT Survey Telescope ATLAS \citep{Shanks15}, as well as deep imaging projects of the regions around nearby massive galaxies such as the Dwarf Galaxy Survey with Amateur Telescopes (DGSAT; \citealt{Martinez10}), the Giant Galaxies, Dwarfs, and Debris Survey (GGADDS; \citealt{Ludwig12}), the Tief Belichtete Galaxien project \citep{Karachentsev15}, and observations taken with the Dragonfly Telephoto Array \citep{vanDokkum14,Abraham14}. Deep imaging programs of nearby clusters, such as the Next Generation Virgo cluster Survey (NGVS; \citealt{Ferrarese12}) and the Next Generation Fornax Survey (NGFS; \citealt{Munoz15}) are also revealing fainter dwarf populations that were previously missed. A large number of dwarf galaxies are expected to be found in these programs, which will greatly contribute to our understanding of dwarf formation and evolution across all environments.

Another deep imaging survey is the Mass Assembly of early Type gaLAxies with their fine Structures (MATLAS) large observing program. MATLAS was designed to study low surface brightness features in the outskirts of nearby massive early type galaxies (ETGs), many of which are isolated, and mapped $\sim$142~deg$^2$ of the sky using MegaCam on the Canada-France-Hawaii Telescope (CFHT), reaching surface brightnesses of $\sim 28.5 - 29.0$~mag/arcsec$^2$  in the \textit{g}-band \citep{Duc15}. These three conditions --- the coverage, depth, and environment of the primary targets --- create an ideal dataset on which to search for low surface brightness dwarf galaxies. Indeed, several tidal dwarf candidates have already been identified in a subset of the images \citep{Duc14}. The MATLAS observations complement previous and ongoing dwarf searches; the MATLAS images are deeper than SDSS ($\mu_g \sim 26.4$~mag/arcsec$^2$; \citealt{York00,Kniazev04}), the imaged galaxies are more distant than those in the ANGST survey, and the targets have early type morphologies unlike other ongoing targeted observing programs.

In this paper, we present a first look at the results of our search for low surface brightness and dwarf galaxies in the MATLAS images. The paper is organised as follows: the MATLAS dataset is described in Section~\ref{section:dataset}, our selection criteria and the construction of the final dwarf catalogue is detailed in Section~\ref{dwarfselection}, in Section \ref{accuracy} we explore potential biases in our sample and test the accuracy and consistency of our classification, while basic properties of the dwarfs are highlighted in Section~\ref{results}, and Section~\ref{conclusions} provides a brief summary of the project.

\section{The MATLAS Dataset}
\label{section:dataset}
MATLAS\footnote{irfu.cea.fr/Projets/matlas/MATLAS/MATLAS.html} is a large observing program conceived under the umbrella of the ATLAS$^{3D}$ legacy survey \citep{Cappellari11}, a program which combines multi-wavelength photometry (radio through optical) with spectroscopic observations and theoretical models in order to study the structural and kinematic properties of a complete sample of early type galaxies in the nearby Universe ($z<0.01$). The ATLAS$^{3D}$ parent sample includes all of the galaxies (both early and late type) that meet the following criteria: a distance between $10 \lesssim d \lesssim 45$~Mpc, a declination $\delta$ such that $|\delta - 29^\circ | < 35^\circ $, a galactic latitude b $ > 15^\circ$, and a K-band absolute magnitude $M_K < -21.5$. 
The first two conditions ensure that the galaxies are visible from the William Herschel Telescope and that key spectral lines would be observable using the Spectroscopic Areal Unit for Research on Optical Nebulae (SAURON) integral field spectrograph, while the latter two requirements were imposed to avoid the plane of the Milky Way and to select a population of massive galaxies. Morphologies were independently classified by members of the ATLAS$^{3D}$ team based on SDSS DR7 images or imaging from the Isaac Newton Telescope to ensure uniformity of the sample. Ultimately, 260 elliptical and lenticular galaxies met the above requirements and were included in the ATLAS$^{3D}$ ETG sample. A more in-depth discussion of the various selection criteria can be found in \citet{Cappellari11}.
 
The MATLAS and the NGVS large observing programs were responsible for the deep optical imaging portion of the ATLAS$^{3D}$ project. NGVS collected data between 2009 and 2013 and mapped 104 deg$^2$ of sky, covering nearly the entire region within the virial radius of the Virgo cluster and the 58 ATLAS$^{3D}$ targets enclosed within, using MegaCam on the CFHT \citep{Ferrarese12}. The MATLAS observations began a year later, with the intention of imaging the remaining ATLAS$^{3D}$ ETGs.

\begin{figure*}
\captionsetup{width=1\textwidth}
\centering
\includegraphics[scale=0.5]{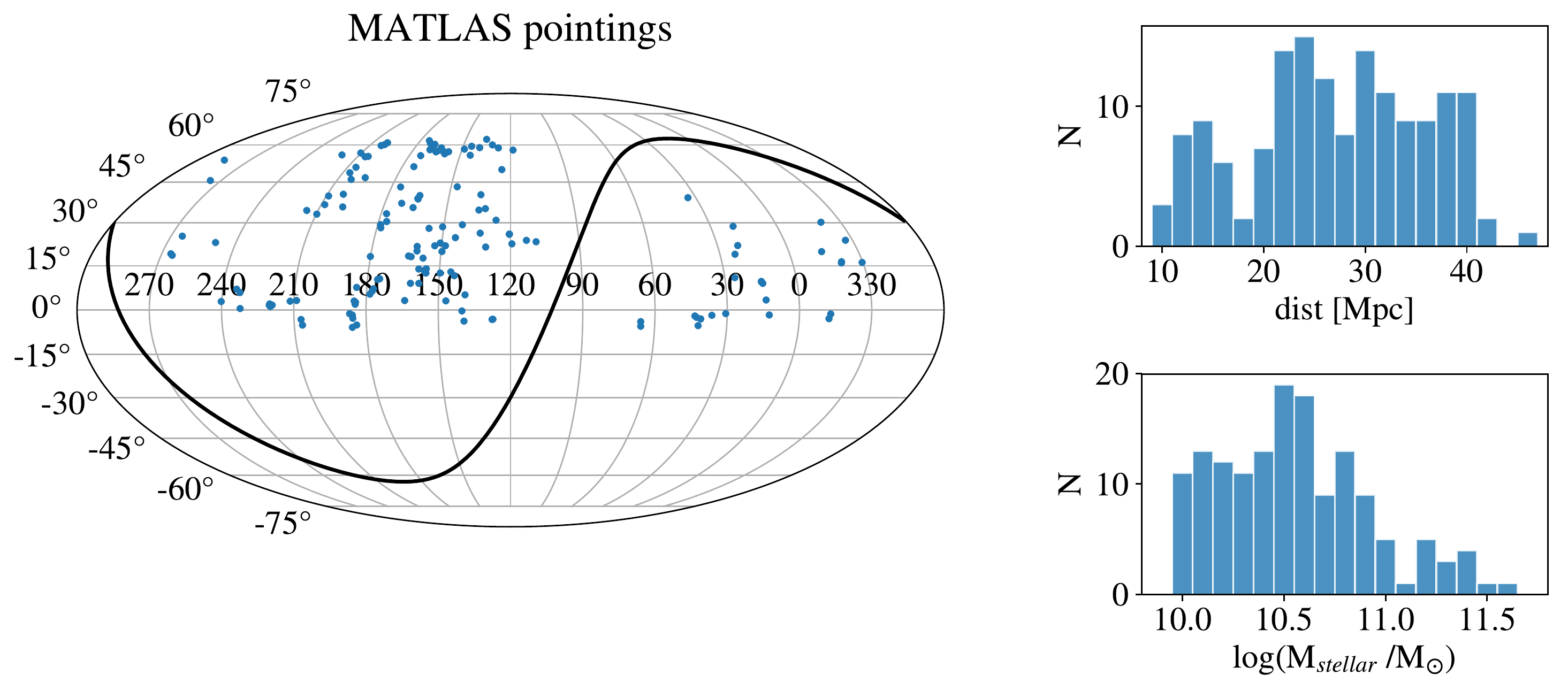}
\caption{Overview of MATLAS observations. \textit{Left:} The equatorial coordinates of the observed MATLAS fields. Each image covers $\sim1.15^{\circ} \times 1.15^{\circ}$ and is smaller than the points. The galactic plane is marked with a solid black line. \textit{Right, top:} A histogram of the distances to the MATLAS targets. \textit{Right, bottom: } A histogram of the stellar masses of the MATLAS targets, calculated from the K-band magnitudes and assuming a mass-to-light ratio $M/L = 1$. Properties of the MATLAS ETGs were taken from \citet{Cappellari11}.}
\label{figure:basics}
\end{figure*}

\subsection{Observations \& Data Reduction}

Observations for the MATLAS project were obtained using MegaCam on the 3.6 meter Canada-France-Hawaii Telescope (CFHT) between 2010 and 2015. The observing strategy, data reduction, and image quality have already been described in detail in \citet{Duc15}, but key details are presented below for completeness. Further details of the MegaCam detector system and data processing can be found in the introductory paper of the NGVS survey \citep{Ferrarese12}.

MATLAS was designed to be a multi-band imaging survey, with complete coverage in \textit{g}, \textit{r}, and \textit{i}-bands with the nearest targets ($d<20$~Mpc) also observed in the \textit{u}-band. The filters have exposure times of $7 \times 345$~seconds in the \textit{g} and \textit{r}-bands, $7\times 230$~seconds in the \textit{i}-band, and $7 \times700$~seconds in the \textit{u}-band\footnote{Due to scheduling problems, some fields have only six exposures while others were mapped twice (12 or 14 exposures). The number of times that specific fields were imaged can be found in Table~\ref{table:MATLAS-targets}.}. Individual exposures were offset varying by 2$\arcmin$~--~14$\arcmin$, in order to utilize the Elixir-LSB pipeline (Cuillandre et al. in preparation). This is an additional reduction pipeline, originally developed for the NGVS, designed to correct for contamination from scattered light in MegaCam. The Elixir-LSB pipeline reduces the global background gradient from $\sim$15\% to $\lesssim 0.4$\% while preserving the 1\% photometric accuracy of the data \citep{Ferrarese12,Duc15}.  

The exposures in each band were run through the AstrOmatic package {\sc{SCAMP}} \citep{Bertin06} to derive a uniform astrometric calibration for each target, followed by the  AstrOmatic package {\sc{SWARP}} \citep{Bertin10} to create the final images. The size of the final image is restricted to the central $\sim63\arcmin\times69\arcmin$ region where the signal-to-noise is highest, and a targeted ETG is located near the centre. The background of the stacked images is flat to approximately 0.2\% of the sky background level, which in turn means that it is possible to detect low surface brightness features almost 7 magnitudes fainter than the sky, locally reaching $\mu_g \sim 28.5$~--~$29$~mag/arcsec$^2$ \citep{Duc15}. Note that the images were calibrated in the MegaCam AB magnitude system.

By the end of the observing program, MATLAS imaged 150 fields in the \textit{g}-band, which encompass 180 ETGs and 59 late type galaxies (LTGs) from the ATLAS$^{3D}$ parent catalogue. Basic properties of the targeted ETGs --- coordinates, distances, and stellar masses --- are illustrated in  
Figure~\ref{figure:basics}. All but one field has data in both the \textit{g} and \textit{r}-bands, while 80 have imaging in \textit{gri} and twelve were targeted in \textit{ugri}. The \textit{g}-band seeing varied from 0.5$\arcsec$ to 1.75$\arcsec$, with an average [median] value of 0.96$\arcsec$ [0.93$\arcsec$]. The images have a resolution of 0.187$\arcsec$/pix, although we used $3\times3$ binned images (final resolution: 0.56$\arcsec$/pixel) for this work to increase the detectability of faint dwarfs. Details of the observations for each field can be found in Table~\ref{table:MATLAS-targets}.

\begin{table}

\centering
\caption{Catalogue of ETGs with MegaCam observations; the first entries can be found below, while the full catalogue can be found online. The columns are: (1) the MATLAS target, (2) the observed bands, (3) the number of exposures for the given filter, (4) the total integration time of all exposures, (5) the image quality; the full width half maximum (FWHM) of the point spread function (PSF), (6) the image background.  Note that a partial version of this catalogue was published in \citet{Duc15}, but the full catalogue is presented here.} 
\label{table:MATLAS-targets} 
\begin{tabular}{l l l l l l  }
\toprule
Galaxy & Band & N & Integration & IQ & Background  \\
& & & [sec] & [arcsec] & [ADUs]  \\
(1) & (2) & (3) & (4) & (5) & (6)  \\
\midrule

IC0560 & \textit{g} & 7 & 2415 & 0.71 & 621.14  \\
& \textit{r} & 7 & 2415 & 0.73 & 994.71  \\
IC0598 & \textit{g} & 7 & 2415& 1.09& 655.00  \\
& \textit{r} & 7 & 2415 &1.00 & 1004.57  \\
IC0676 & \textit{g} & 7 & 2415 & 1.14& 652.86  \\
& \textit{r} & 7 & 2415 & 0.41 & 886.86  \\
IC1024 & \textit{g} & 7  &2415  & 1.08 &   707.71\\
& \textit{r} & 7 & 2415 & 1.16 & 882.71 \\
NGC0448& \textit{g} & 7 & 2415 & 0.64& 704.29  \\
& \textit{r} & 7 & 2415 & 0.77 & 941.57  \\
& \textit{i} & 7 & 2415 & 0.65&1054.86  \\

\bottomrule
\end{tabular}
\end{table}

\section{Dwarf Candidate Selection}
\label{dwarfselection}
There are three typical approaches to identifying low surface brightness galaxies: identification through a visual inspection of each image, a semi-automatic approach where an automated catalogue is cleaned visually, or generating a fully automated catalogue. The first two methods are most commonly used; both the NGFS and GGADDS teams visually examined their fields to identify potential dwarfs, while the NGVS, DGSAT, and Dragonfly projects have used semi-automated catalogues. Software such as the Markovian Software for Image Analysis in Astronomy program MARSIAA \citep{Vollmer13}, NoiseChisel \citep{Akhlaghi15}, ProFound \citep{Robotham18}, and DeepScan \citep{Prole18} may be capable of generating more precise automated lists of low surface brightness dwarf galaxies, but are not yet in wide use.  Due to concerns about the completeness and purity of the generated catalogues, we did not pursue a fully automated catalogue, but instead created both a visual and a semi-automated list of dwarf candidates.

\subsection{Visual Catalogue}

The visual catalogue (hereafter the `full-visual catalogue' to distinguish it from the galaxies identified in the semi-automated dwarf catalogue) was created by a simple visual inspection of the MATLAS images. It should be noted that this catalogue was never intended as a stand-alone product. Rather, these galaxies were used to estimate the detection success rate of {\sc{Source Extractor}} (hereafter SExtractor; \citealt{SExtractor}) and to test the best selection criteria to use when generating the automated candidate list. 

The full-visual catalogue was constructed in a two step process. All team members involved in the classification first searched five training fields  (NGC~448, NGC~2685, NGC~4382, NGC~5308, and IC~1024) for extended, low surface brightness galaxies. This was done to familiarize ourselves with the images, learn to distinguish dwarfs from other features such as cirrus, identify and correct biases of individual team members, and provide an internal check on our consistency. At this time we also agreed to conservatively reject dwarf candidates with diffuse halos but extended central light concentrations --- which are easily confused with background lenticular galaxies --- and any galaxy exhibiting spiral structure. After compiling the individual catalogues from the five training fields, we identified 74 dwarf candidates. The visual search for dwarf candidates was then extended to all fields, with each field inspected by at least one team member. This yielded a final sample of 1349 candidates, after removing duplicates from overlapping fields.

\subsection{Semi-automated dwarf catalogue}
\begin{figure}
\includegraphics[width=0.47\textwidth]{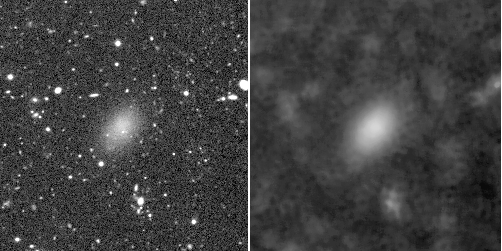}
\caption{An example of one visually identified dwarf on the \textit{g}-band image and the same object in the {\sc{rmedian}} filtered image. Each cutout is 2.35$\arcmin \times$2.35$\arcmin$.   }
\label{figure:dwarf_separation0}
\end{figure}

\begin{figure*}
\includegraphics[width=\textwidth]{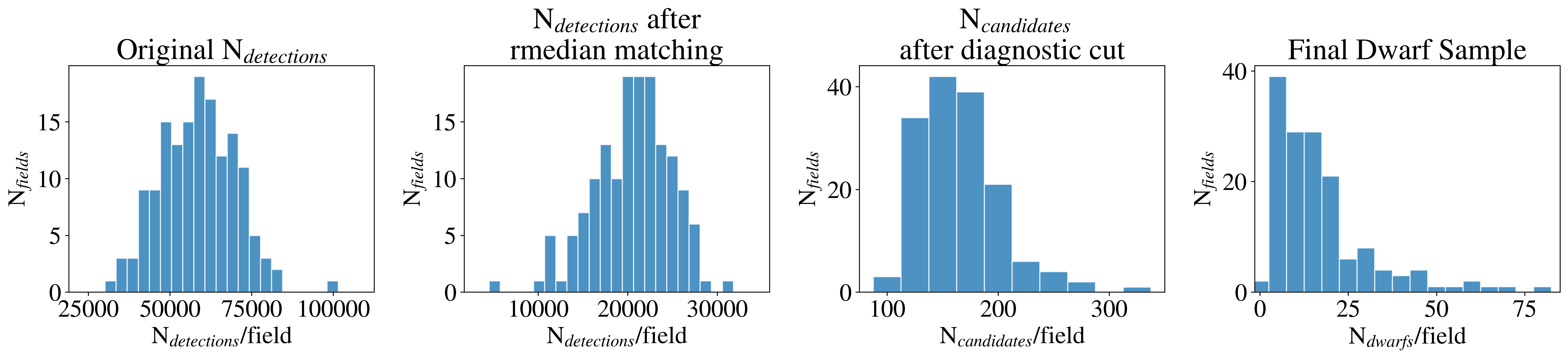}
\caption{The number of potential dwarfs at key stages of the catalogue cleaning process. \textit{From left to right:} (1) The number of SExtractor detections on the \textit{g}-band images per field. (2) The number of detections per field after matching against the {\sc{rmedian}} filtered images. (3) The final number of dwarf candidates after applying the vertical adjustments and applying an area cut. (4) The final number of dwarfs identified per field; candidates that lie within the bounds of more than one image are included in the dwarf count for each field in which they appear. }
\label{figure:dwarf_separation}
\end{figure*}

An automated list of dwarf candidates was generated using output parameters from SExtractor. However, SExtractor was not designed explicitly for the detection of low surface brightness features. In order to enhance the faint signals of these sources and increase the likelihood of detection, we first applied a ring median filter on each \textit{g}-band image using the {\sc{IRAF}} software package {\sc{rmedian}}. This technique, presented in \citet{Secker95}, replaces the flux at each pixel by the  median value from a user defined annulus around that point. Consequently, objects with diameters less than $r_{inner}$ will be removed from the image and extended low surface brightness objects are enhanced. An example of the output from {\sc{rmedian}} is shown in Figure~\ref{figure:dwarf_separation0}.  The size of the annulus was deliberately kept small, with $r_{inner} = 3$~pix and $r_{outer} = 8$~pix, to ensure the detection of the smallest dwarf candidates identified in the visual sample. SExtractor was then run on both the original \textit{g}-band image and the {\sc{rmedian}} filtered images. We required the sources to be detected in both images, but did not impose any detection criteria on the other filters, as not all fields were observed in the other bands. The filtering procedure reduced the number of detections from 8,922,424 on the original images (using a 1.5$\sigma$ detection threshold) to 3,120,066 sources, or $\sim$35\% of the original detections. Per field, the number of matched detections varies from 5,335 to 30,948 objects. Histograms illustrating the number of sources ($N$) per field, at each of the major stages of catalogue cleaning, are shown in Figure~\ref{figure:dwarf_separation}.

We tested several parameters from SExtractor to identify the best method of isolating low surface brightness galaxies from background objects: surface brightness (both central and average), apparent magnitude, compactness, and various size estimates. Initially, we also tested parameters from {\sc{Galfit}} \citep{Galfit1,Galfit2}, such as the effective radius ($R_{e}$) and S{\'e}rsic Index ($n$), but it was not feasible to model the $>3\times10^6$ sources in the {\sc{rmedian}} matched catalogue in a reasonable time. The visually identified dwarf candidates were used to identify the location of the dwarfs in each parameter space and test the separation of dwarf and non-dwarf galaxies. 

\begin{figure}
\includegraphics[width=0.47\textwidth]{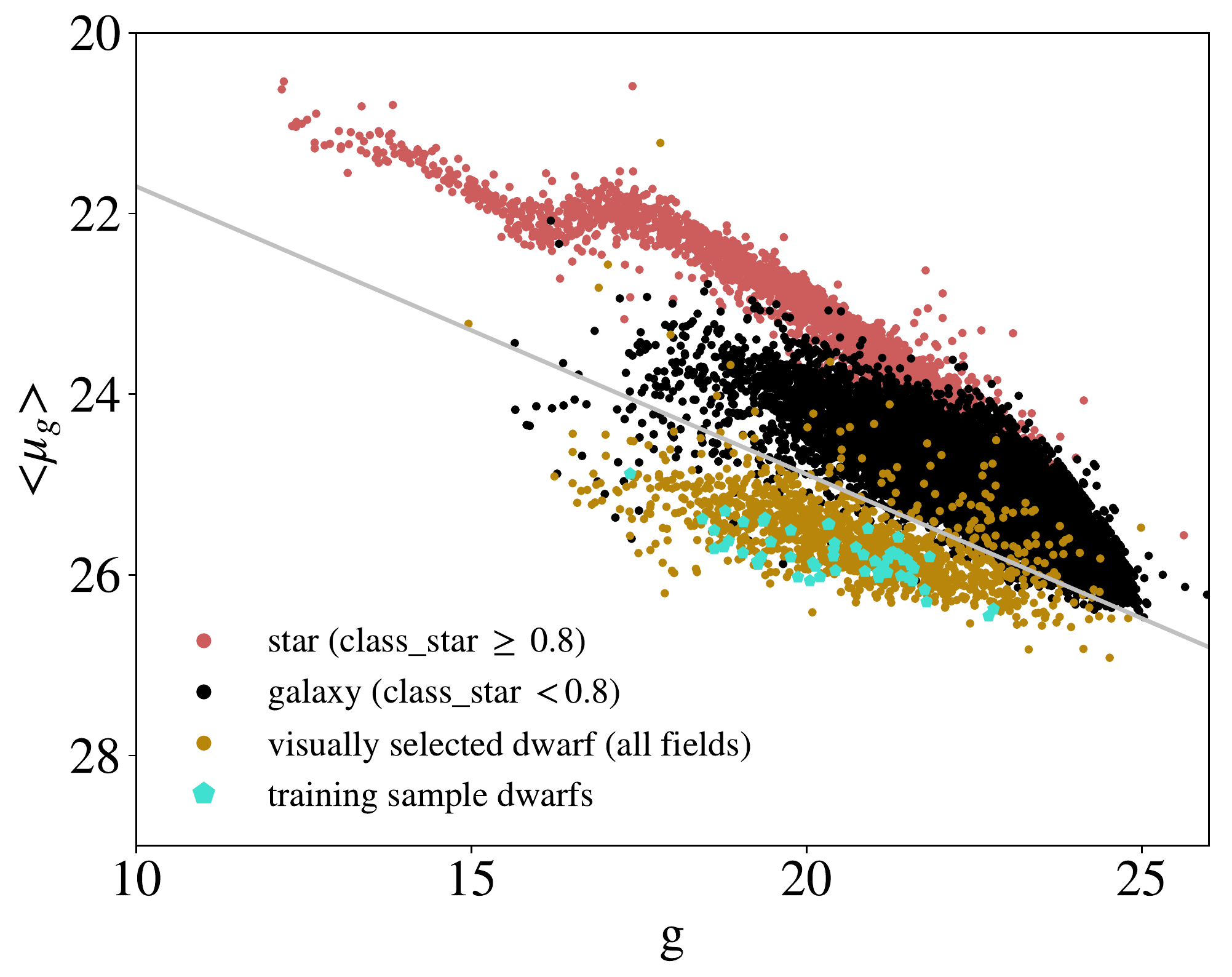}
\caption{The separation of dwarf and non-dwarf galaxies in $\langle \mu_g \rangle$ vs.\ $m_g$ space. The turquoise pentagons are the dwarf candidates identified by multiple authors in the five training fields, while the gold circles were flagged as dwarf candidates by at least one team member in a visual inspection of all fields. See text for a discussion of the scatter. The galaxies (black) and stars (red) show all of the SExtractor detections from one representative field (NGC~4382); these two populations were separated using the SExtractor class\_star parameter, which returns the probability that a source is point like or extended. The dwarf and non-dwarf candidates appear to follow two distinct relations, and the dwarf candidates were selected using a linear cut like the grey line in all fields.    }
\label{figure:diagnostic}
\end{figure}

The cleanest separation was found in plots of average surface brightness ($\langle \mu_g \rangle$) versus apparent magnitude ($m_g$), which is illustrated in Figure~\ref{figure:diagnostic} using the dwarfs from the training sample (turquoise), full-visual sample (gold), and all SExtractor detections from a representative field, NGC~4382. Average surface brightnesses were chosen instead of the more common central surface brightness ($ \mu_{0,g}$) as this slightly reduced the scatter observed in the visually identified dwarfs, and we did not convert $m_g$ to absolute magnitudes to avoid introducing biases into the data by assuming a distance for all galaxies in the images. 

The dwarf and non-dwarf populations appear to follow two relations in Figure~\ref{figure:diagnostic}, but the separation is not perfect. Roughly 10\% (142) of the dwarfs in the visual catalogue are scattered into the locus of background galaxies; some of these correspond to poor SExtractor extractions --- either due to fragmented galaxies or dwarf candidates that were not properly deblended from nearby sources --- but others may be background contaminants in the full-visual sample as this catalogue contains galaxies flagged by a single person. None of the dwarfs in the training sample, which was more rigorously inspected, lie in the locus of background galaxies. Additionally, there are SExtractor detections that fall in the dwarf region of interest, which were not visually identified as dwarf candidates; galaxies at the bright end are typically massive spiral or lenticular galaxies, while false detections of ghost halos and cirrus fall throughout the region populated by the dwarfs. 

A simple linear cut, roughly tracing the upper edge of the visually identified dwarf population but positioned next to the lower boundary of the main locus of galaxies, was used to select potential dwarf candidates in each field. Due to small shifts in the position of the background galaxies, likely due to variations in extinction corrections across the fields (not corrected for at this stage) and variations in the distribution of background galaxies included in the images, minor adjustments were applied to the diagnostic line in each field; most fields required less than a 0.1~mag/arcsec$^2$ vertical offset, and the diagnostic diagram of each field was visually inspected to verify that the selection cut was positioned correctly. An example of the selection cut used in field NGC~4382 is shown in Figure~\ref{figure:diagnostic}. With this cut, we identified 46,287 potential dwarf candidates over all 150 MegaCam fields.

The slope of the diagnostic line is slightly shallower than the lower edge of the main locus of galaxies, and the two intersect at the faint end. Almost half of the galaxies selected by the diagnostic cut are small galaxies from this region which were not removed when matching against the {\sc{rmedian}} filtered images. A subsample of these small galaxies were binned by size (using the \textit{isoarea} parameter from SExtractor), and a visual inspection determined that we could identify structures in background galaxies only when the area of the object was larger than $\sim75$~pix$^2$ ($\sim 23.5$~arcsec$^2$). Because we would not be able to accurately distinguish smaller galaxies as nearby dwarfs from distant background galaxies, we dropped all galaxies with an isoarea smaller than 75~pix$^2$ from the sample. For comparison, such a cut would allow us to retain dwarfs with physical sizes similar to Fornax and Sculptor to distances of approximately 45~Mpc and 63~Mpc, respectively. This area cut reduces the number of dwarf candidates to 25,254 objects over the entire sample, with individual fields containing anywhere from 97 to 314 dwarf candidates. 

Our catalogue of visual dwarfs was then matched against the final automated list of candidates from SExtractor, and any galaxy that was not included in the automated catalogue was manually added to the sample. In total, there were 268 additions; 142 candidates (10.5\% of the full-visual catalogue)  were originally excluded by our selection criteria and 126 (9.3\%) were not detected by SExtractor. Thus, the final catalogue of potential dwarfs contained 25,522 galaxies. 

\subsubsection{Cleaning the automated catalogue}
The automated catalogue was cleaned by nine of the authors working in parallel, and each dwarf was visually inspected by at least three team members using an online interface that was designed for this task. The visual classifications were based on a series of five images at different stretches (one of which was a composite true colour image) for each candidate which were created using {\sc{fitscut}} and {\sc{stiff}} \citep{Bertin12}; each cutout was centred on the candidate and had fixed dimensions ($\sim$1$\arcmin \times$1$\arcmin$) and fixed dynamical ranges so the size and relative brightnesses of the candidates were meaningful.  Team members were able to rank the likelihood of the candidate being a dwarf, with responses including `yes', `likely', `unsure', and `no'. Any galaxy that was given at least one yes/likely vote or two or more unsure responses was extracted as an object of interest, resulting in a final catalogue of 3644 potential dwarf candidates; this number drops to 3311 once the duplicates are removed (see Section~\ref{section:consistency} for a discussion of our classification consistency of the objects in overlapping adjacent fields). 

While classifying the candidates, we also assigned morphologies to all of the dwarf/likely dwarf candidates in the sample. Dwarfs were separated into four broad categories: `dE/dSph', `dIrr/BCD', `compact dwarf', and `unsure'. The final morphological class was assigned to the candidates based on the majority opinion of the classifiers. In the event that there was an even split, e.g. two team members classified the object as a dwarf but assigned different morphological classifications or all classifiers marked the candidate as an `unsure' dwarf, the galaxy was given an unsure morphology. Overall, there was good agreement in the morphological classifications, with fewer than 10\% given discrepant morphological classifications. 

\subsubsection{The final dwarf catalogue}
To ensure the uniformity of the sample and remove any suspected background interlopers remaining in the sample, five authors performed one final review of all the galaxies in the preliminary catalogue. In addition to the image cutouts, the classifiers were also presented with spectroscopic redshifts --- when available --- from SDSS DR13 \citep{Blanton17,Albareti17} as well as {\sc{Galfit}} model parameters (e.g., $R_{e}$, $n$, $M_g$; to calculate $M_g$, we assumed that the dwarfs were at the same distances as the targeted ETG in each field) along with the  model residuals.   {\sc{Galfit}} was initially run on just the dE subsample from the preliminary catalogue, and only those values were available for this secondary review, but the modelling has since been expanded to the full sample and the results will be presented in a future paper (Poulain et al.\, in preparation). In brief, {\sc{Galfit}} was run iteratively on each candidate to obtain a clean model for the galaxy using a single S{\'e}rsic profile. The first application used parameters returned by SExtractor as the initial inputs, after which sources at distances  $> 1 R_{e} $ which were identified in the SExtractor segmentation image were  replaced using the modelled flux (plus noise) at those pixels. In addition, bright point sources on the dwarf candidates ($d \leq 1 R_{e} $) were identified using {\sc{DAOPHOT}} \citep{Stetson87} and removed the same way.  {\sc{Galfit}} was then run a second time on the cleaned image. Any point-like sources remaining within $\sim4\arcsec$ of the photometric centre in the final model residuals were flagged as potential nuclei. For this work, the nuclei have been masked and the galaxies were re-modelled to obtain parameters for only the diffuse component of the dwarfs.

Our final dwarf selection is based on the majority opinion of the five authors who participated in the final review. Ultimately, we were left with 2210 unique dwarf candidates. Dwarf candidates were identified in all but one image; we attribute the lack of dwarfs in field NGC~3499 to the lower quality of the image, which was badly contaminated by light from a nearby star, rather than an actual absence of dwarfs around that galaxy. In the remaining 149 fields, we identified between two and 79 dwarf candidates per image, with a median of 17 candidates per field. The distribution of $N_{dwarfs}$/field can be found in Figure~\ref{figure:dwarf_separation}, \textit{right}. Dwarfs that fall within the bounds of multiple, overlapping images are included in the dwarf count for each field.   
 
\begin{figure}
\centering
\includegraphics[width = 0.47\textwidth]{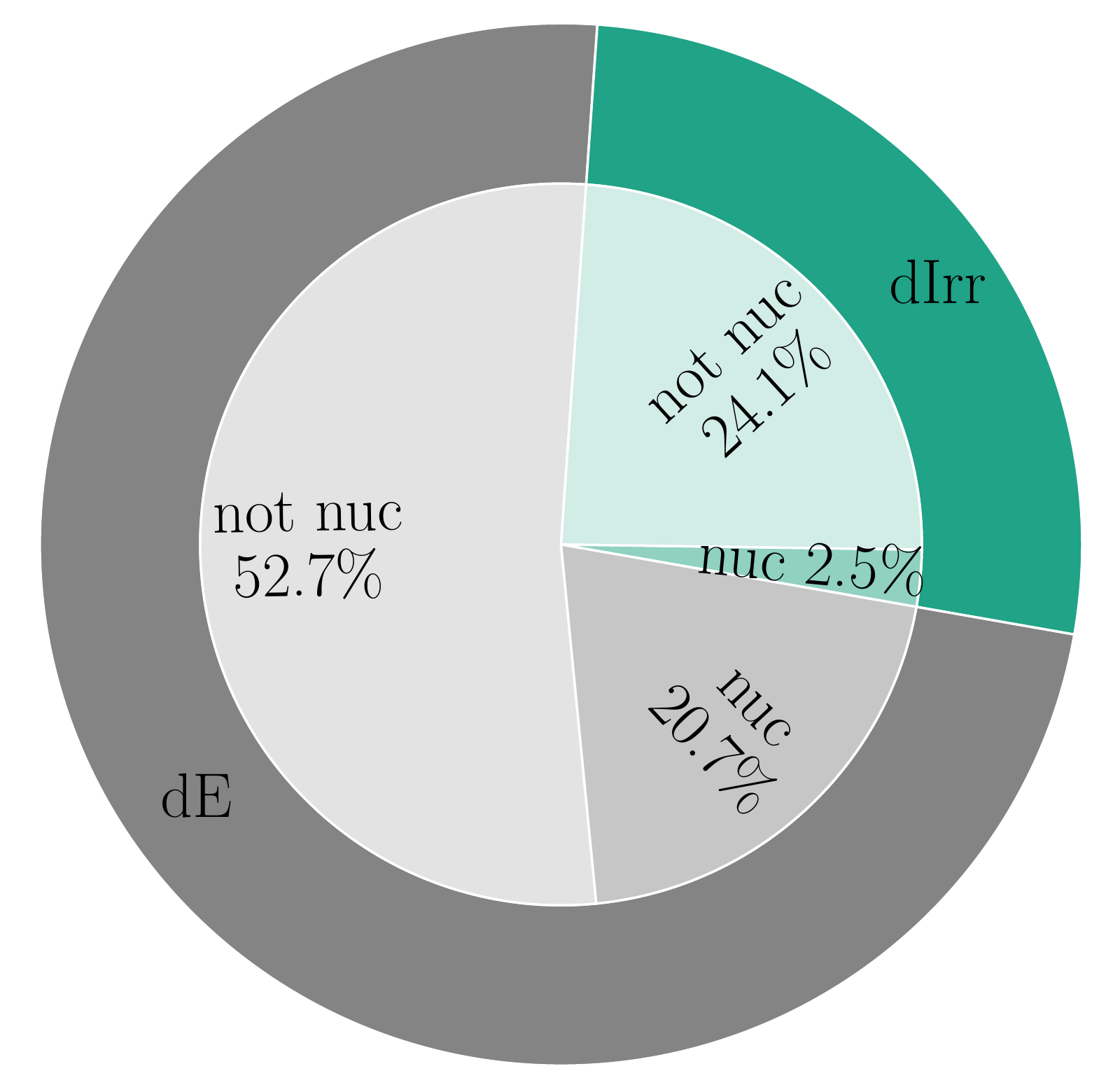}
\caption{The final morphological classifications of our 2210 dwarf candidates. The morphology is shown in the outer ring, while the inner wedges show the breakdown of nucleated/non-nucleated candidates within each class. The percentages are given with respect to the full sample.   }
\label{figure:finalmorphologies}
\end{figure}

\begin{figure*}
\centering
\includegraphics[width = 0.95\textwidth]{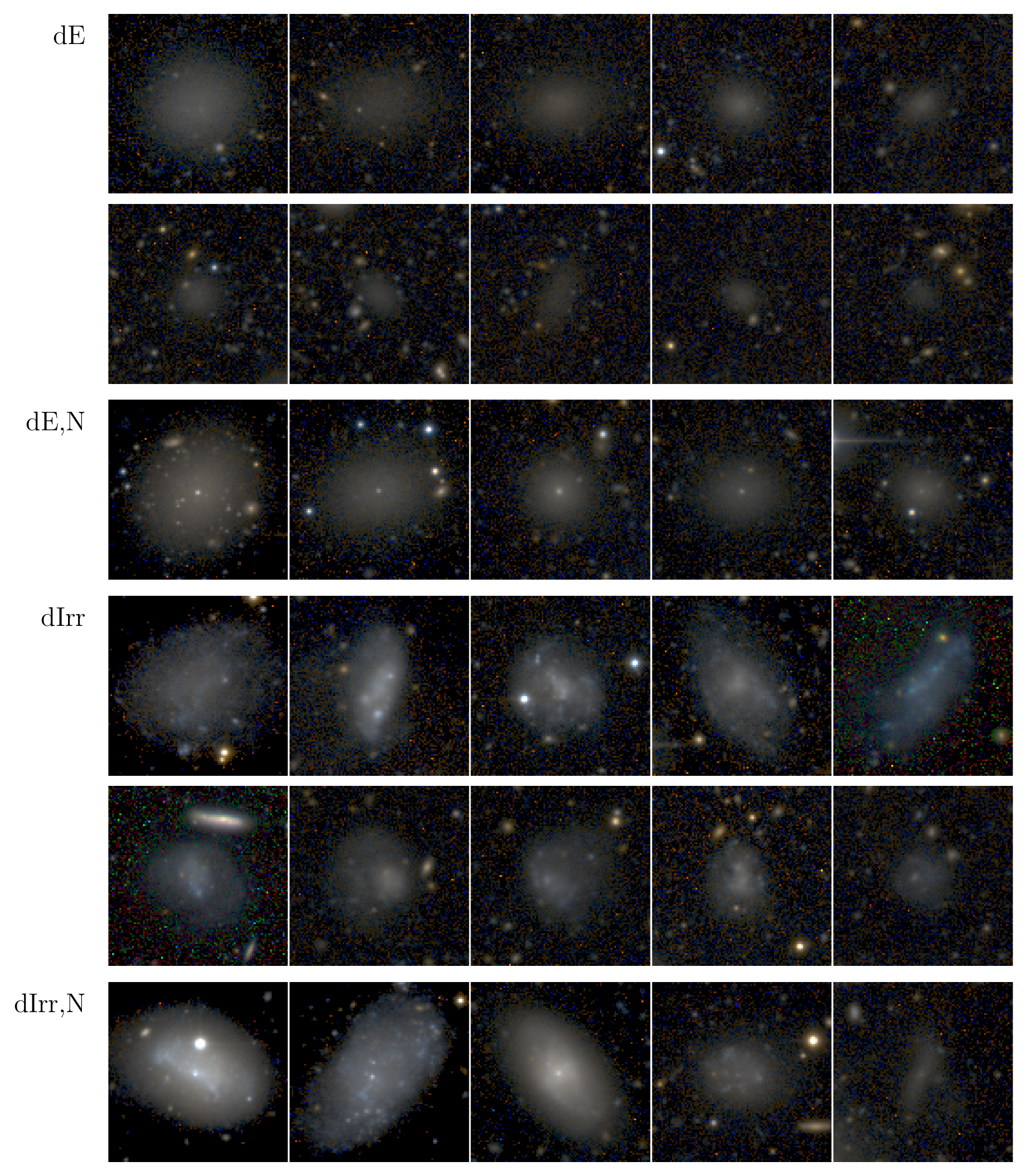}
\caption{Examples of dwarf candidates in our final dwarf sample, sorted by morphology and apparent magnitude.  Each cutout is $\sim 1 \arcmin \times 1 \arcmin$ (this corresponds to $\sim3$~kpc on a side at a distance of 10~Mpc or 13~kpc on a side at 45~Mpc), and the colour composites were created using {\sc{stiff}}. }
\label{figure:finalimagesexamples}
\end{figure*}

In the final review, the classifiers were also asked to confirm the morphological classifications of the candidates, and galaxies with unsure morphologies were classified. Our sample is dominated by dE candidates (73.4\%), while 26.6\% were classified as dIrrs. Only 23.2\% of the candidates are nucleated, most of which are dE,N candidates as opposed to dIrr,N. The morphological breakdown is illustrated in Figure~\ref{figure:finalmorphologies}, and examples of each classification from our final catalogue are shown in Figure~\ref{figure:finalimagesexamples}.

\section{The accuracy and consistency of our dwarf classification}
\label{accuracy}

\subsection{Biases}
\label{biases} 
\begin{figure}
\centering
\includegraphics[width = 0.48\textwidth]{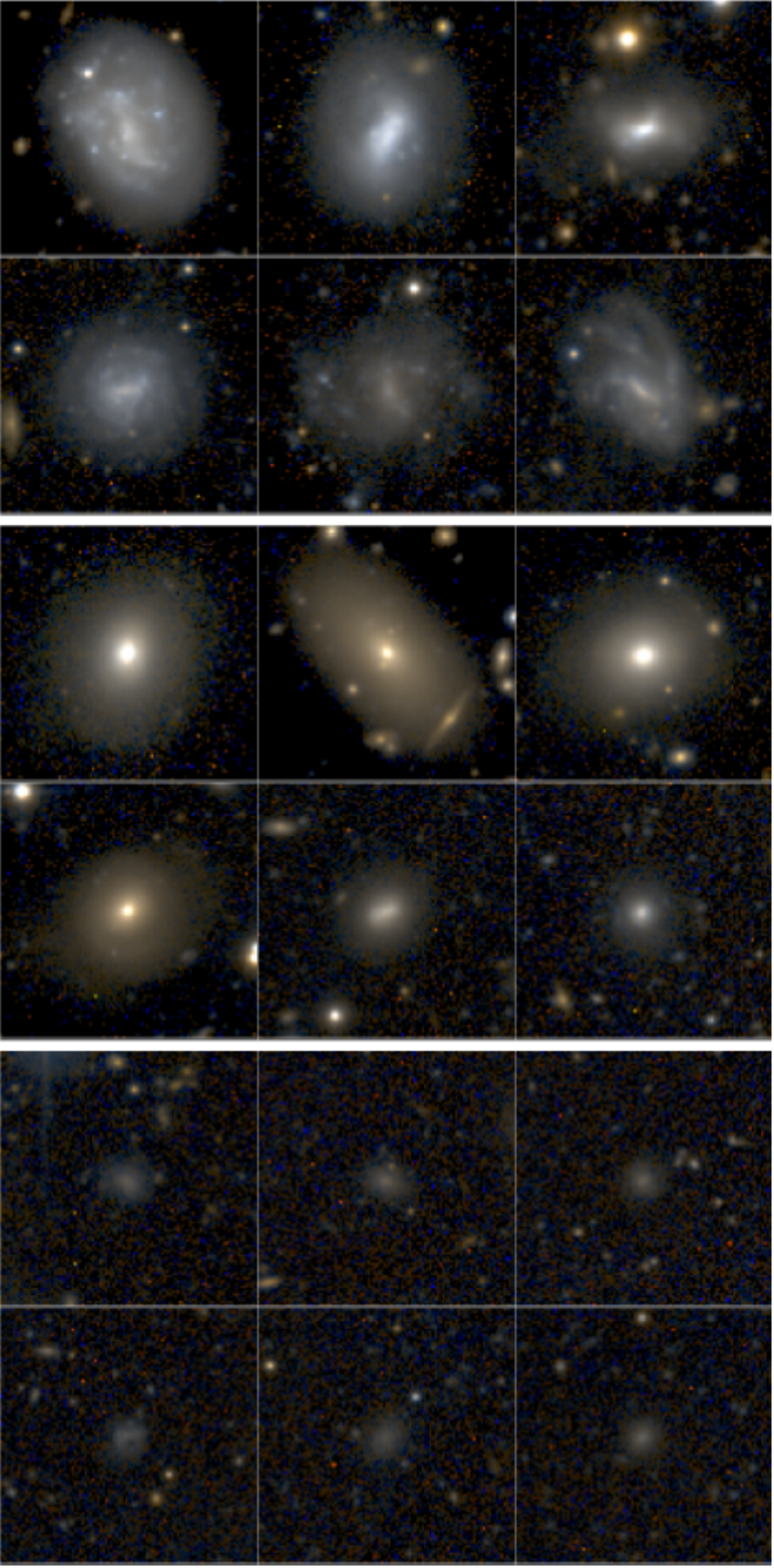}
\caption{Examples of galaxies that were rejected as dwarf candidates. Galaxies in the top two rows were considered to have too much structure, galaxies in the middle two rows had too large nuclei, and the galaxies in the bottom two rows were considered too small and/or too difficult to separate from background galaxies. Each cutout is $\sim 1\arcmin \times 1\arcmin$ (this corresponds to $\sim3$~kpc on a side at a distance of 10~Mpc or 13~kpc on a side at 45~Mpc), and the colour composites were created using {\sc{stiff}}. }
\label{figure:rejected}
\end{figure}

Biases are built into any visually identified sample of galaxies, but quantifying them is not simple as we have no way to estimate the number of genuine dwarf galaxies that were rejected. However, there are four galaxy types that we acknowledge were more difficult to classify consistently: (1) galaxies with diffuse halos but more extended central light concentrations, (2) dIrr galaxies, (3) small dwarf candidates, and (4) bright/large candidates. Galaxies in the first category are easily confused with background lenticular galaxies, thus we decided to conservatively reject them. However, the size of observed nuclear emission appears to vary as a nearly continuous function, and it was not always obvious how to clearly discriminate between background galaxies and nearby dwarfs. Similarly, there was also some confusion when separating dIrr from galaxies with coherent structures that may have been highly disturbed spirals; although some dwarf spiral candidates have been identified in the literature (see, for example, \citealt{HidalgoGamez04}), most are ultimately found to be more distant, massive spirals and the team collectively agreed to reject any galaxy with spiral structure. This potentially led to some dIrr candidates being rejected. In addition, there is some remaining uncertainty in the classification of the smallest candidates, even after applying the area cut to the preliminary sample, and the depth of the MATLAS imaging could conversely make some of the larger dwarf candidates appear too bright in the cutouts. Examples of galaxies in the first three categories that were included in the preliminary sample but rejected in the final sample are shown in Figure~\ref{figure:rejected}.

Although it is not possible to estimate the number of dwarfs within each category that we have rejected from the preliminary catalogue, we can explore potential biases in the sample by examining the confidence level of the dwarf classification in the preliminary catalogue as well as the overall fraction of various dwarf populations that we kept in the final catalogue. The objects of interest in the preliminary catalogue can be separated into two broad classes:  {\textit{confident}} dwarfs which received two or more yes/likely classifications and and {\textit{less confident}} dwarfs (the remaining candidates). Of the 654 dIrr candidates in the original catalogue, 397 ($\sim61$\%) are ranked as less confident dwarfs. During the final cleaning, we rejected 30\% of the dIrr candidates, but only 10\% of the dE candidates. These numbers can be interpreted two ways: either a larger fraction of the dIrr candidates in the preliminary catalogue are questionable (possible, given the confusion with very distorted spirals) and should have been dropped, or we were more conservative regarding the dIrr candidates in the secondary review and rejected good dIrr galaxies due to user bias. Without distances, it is difficult to determine which scenario is more likely. Nor can we constrain the fraction of dIrr that may have been missed in the preliminary catalogue. 

The nucleated and non-nucleated populations do not show such a discrepancy, although there was potential confusion between nucleated dE galaxies and massive lenticulars. For the nucleated dwarfs, $\sim65$\% were rated confident dwarfs. In addition, the fraction of nucleated dwarfs is similar in the preliminary and final catalogues (18.5\% and 20.5\%, respectively), suggesting that we were at least not biased against the nucleated dwarfs.

It should also be noted that we do not expect compact dwarfs in the sample, and no Ultra Compact Dwarfs (UCDs). An UCD with $R_{e} = 100 $~pc has an angular radius of $\sim2\arcsec$ at a distance of 10~Mpc; this would give the largest UCDs an area of $\sim$50 pix$^2$, assuming it is perfectly circular, which would have been cut from our sample.

Biases can also be introduced by the detection software. For example, dwarf candidates that lie within the extended halo of a massive galaxy or star were unlikely to be correctly deblended and extracted by SExtractor. We can place constraints on this, however, using the visual catalogue of dwarfs. There were 139 candidates identified by eye that were not detected by SExtractor, roughly 10\% of the visual sample.

\subsection{Classification of duplicate galaxies}
\label{section:consistency}
While classifying the candidates, we implemented several checks to test our consistency. For example, some fields overlap and we intentionally kept any duplicated objects in the semi-automated catalogue to test how these objects were classified by different team members. There are 1187 candidates that appear more than once in the full catalogue of 25,522 objects. 

Rather than comparing individual classifications directly, which would unnecessarily magnify minor deviations, such as one team member classifying the candidate as a dwarf and another selecting `likely dwarf', we again separate the duplicated objects into the confidence classes defined above, and add a {\textit{rejected}} class for those that were not included in the preliminary catalogue. Using these three categories, 84.6\% of the duplicated objects were classified consistently. A further 75 galaxies (5.9\%) were consistently retained in the preliminary catalogue but were alternately classified as confident or less confident candidates. The subdivisions between these classes are somewhat arbitrary, so this deviation is not particularly worrisome. Of more concern are the remaining galaxies which were alternately included in or rejected from the preliminary catalogue; 8.3\% were either rejected or classified as less confident dwarfs while 1.2\% of the galaxies were classified as confident dwarfs by one set of classifiers and rejected by another. Some of these discrepancies are justified; the dithering pattern and image stacking used by MATLAS results in images with uneven coverage near the edges, which may cause faint candidates near the edge of any field to be lost due to lower signals. In addition, at least one duplicate was bisected by the edge of the image, making it difficult to classify.

\subsection{Comparison with other imaging programs}

We can directly compare our dwarf selection methodology against the DGSAT and Dragonfly projects, as the three observing programs have a slight overlap. Seven candidates in these two programs fall within the MATLAS footprint: NGC5457-DF-4, NGC5457-DF-6, and NGC5457-DF-7 were independently identified as dwarfs by \citet{Merritt14} and \citet{Javanmardi16}, while \citet{Henkel17} discovered the dwarf candidates NGC3625-DGSAT-1, NGC3625-DGSAT-2, NGC3625-DGSAT-3, NGC3625-DGSAT-4.  Four of the seven galaxies were detected by SExtractor, and one additional candidate (NGC5457-DF-4) was identified in the full-visual catalogue and subsequently appended to the automated list; the two faintest targets ($\mu_{0,g}\geq 27.5$~mag/arcsec$^2$; \citealt{Merritt14}) can be found in the MATLAS imaging but they are so faint that they were never selected as potential dwarf candidates. Of the five galaxies we classified via the online interface, two were rejected during the first classification round --- one due to its overall faintness and asymmetric structure, and the other for having too much structure for its size --- while the remaining three are included in our final catalogue. This illustrates not only the difficulties in visually selecting low surface brightness galaxies, but the importance of image depth and resolution in their identification.  

One of our fields, NGC~4382, lies at the edge of the Virgo cluster and overlaps with an earlier study of dwarfs in the Virgo cluster by \citet{Trentham02}, who probed the luminosity function of galaxies in the cluster with absolute magnitudes between $-22\leq M_B \leq -10$. Their imaging is not as deep as the MATLAS survey, but this still provides an important check on our consistency among brighter dwarf candidates. Trentham \& Hodgkin identified ten dwarf galaxies that lie within the image bounds of the MATLAS field. Eight of these galaxies were included in the preliminary catalogue (one was not properly deblended from the diffuse halo around NGC~4382 and therefore was never detected by SExtractor, while the other dwarf was lost when matching against the ring-median filtered images) and all eight were retained in our final catalogue. 

Another catalogue, the CVCG \citep{Ann15}, contains $\sim5800$ galaxies in the nearby Universe, of which 2623 were classified as dwarfs (according to their assigned morphologies). Of these, 304 fall within the MATLAS images. However, only 201 of these galaxies were in the automated catalogue, and only four additional galaxies were flagged as dwarfs in the full-visual catalogue. There are a number of reasons why these galaxies were missed: some are bisected by the edges of the MATLAS fields, are located near image defects, or are found in regions with higher local background levels which may have led to problems deblending the dwarf from the surrounding emission. A number of the CVCG dwarfs were also found to have extended central light concentrations and were then excluded by our selection criteria; since we decided to reject such galaxies due to potential confusion with background S0 galaxies, these dwarfs were also excluded from our full-visual catalogue.  

If we consider only the CVCG candidates within our preliminary catalogue, the agreement is much better. We retained 176 (83\%) of the known dwarfs in our final catalogue.  The CVCG catalogue also contains morphological classifications of all the dwarf candidates, which can be used to test our morphologies, although they applied finer subdivisions than we did. If we group all of their dE, dS0, and dSph candidates under the umbrella of `dEs' and merge the BCDs with the dI classifications, we find an 87\% agreement between our morphological classifications.

\subsection{Probing the nature of our dwarf candidates with distance measurements}
\label{bfcontamination}
Perhaps the greatest uncertainty in this work is the distance to the galaxies within our sample, without which we cannot confirm the dwarf nature of the candidates or the physical association between the dwarf candidates and the massive ETGs/LTGs in the images. To combat this problem and estimate the level of background contamination, we cross matched our list of dwarf candidates against the SDSS DR13 database to search for objects with measured spectroscopic redshifts.  We further required a median S/N$\geq 5$ for the spectra; galaxies with lower signal-to-noise ratios were dropped to avoid contamination from bad $z_{spec}$ fits. From our initial catalogue of 25522 objects, there were 3984 galaxies that met these criteria, 122 of which are in our final dwarf catalogue. Spectroscopic redshifts were also obtained for 218 candidates from the CVCG, which was compiled from sources in the Korea Institute for Advanced Study Value-Added Galaxy Catalogue (KIAS-VAGC; \citealt{Choi10}) which is based on SDSS data, the NASA Extragalactic Database (NED), and nearby galaxy groups identified by \citet{Makarov11}.

The spectroscopic redshifts were also augmented with HI detections from the ATLAS$^{3D}$ project \citep{Serra12} and the ALFALFA Extragalactic HI Source Catalogue \citep{Haynes18}. Serra et al.\ compiled HI data,  using a combination of archival data and their own observations with the Westerbork Synthesis Radio Telescope (WSRT) for the regions around 166 ETGs in the ATLAS$^{3D}$ survey (127 of which lie outside of the Virgo cluster). The observations used a bandwidth of 20~MHz sampled in 1024 channels, which corresponds to helio-centric velocities in the range $v_{helio}\lesssim4000$~km/s. The reduced datacubes have a velocity resolution of 8.2~km/s after Hanning smoothing. The datacubes were searched within a 1$\arcmin$ radius (an average beam size for the declinations probed) of the candidates; peaks were fit with a Gaussian profile and the velocity of the candidate was extracted from the midpoint of the half-maximum. Forty-nine candidates have a corresponding HI detection in the WSRT data, which was confirmed by a follow-up visual inspection. 

The ALFALFA HI survey was designed as a `blind' survey using the Arecibo telescope, which mapped $\sim7000$~deg$^2$ of the sky using a drift scan technique \citep{Giovanelli05}, and at each position the full frequency range 1335--1435~MHz ($-2000 <v_{helio}<18000$~km/s) was probed; the channel width is 5.1~km/s. The ALFALFA Extragalactic HI Source Catalogue contains 31500 HI sources at distances $d \lesssim 100$~Mpc and provides recessional velocities for the galaxies, as well as HI mass estimates and rotational velocities. The catalogue was matched against our final dwarf sample using a search radius of 4$\arcmin$, the average beamsize of the telescope, and the association of the detections with the dwarf was again confirmed through a follow-up visual confirmation on the MATLAS images. There are 57 dwarf candidates with a match in ALFALFA and a $S/N \geq 5.0$.

Finally, we also matched our final dwarf candidates against the Updated Nearby Galaxy Catalogue (NEARGALCAT; \citealt{Karachentsev13}) and the catalogue of Local Group dwarfs \citep{McConnachie12} to check for foreground contaminants. No Local Group dwarfs fall within the MATLAS images, but there were 23 matches in the NEARGALCAT. It should be noted that distances reported in the catalogue were determined using various distance indicators, and may not agree with the distances calculated from $z_{spec}$ or $v_{HI}$ from the other samples. 

Several of the candidates have multiple distance indicators; ultimately there are 292 candidates ($\sim13$\% of our final sample) with at least one measurement from the above sources. These candidates  are split evenly by morphological type and include 142 dE and 150 dIrr candidates. Ignoring any peculiar velocities and converting the various measurements to distances via $d = v/H_0 = cz/H_0$, we find distances for the dwarf candidates between 5.1 and 100.6~Mpc. The median distance of this sample is 26.5~Mpc and roughly 90\% of the dwarf candidates (263 galaxies) lie within the 10~--~45~Mpc range populated by the MATLAS ETGs. Eleven of the remaining dwarf candidates lie at distances $<10$~Mpc (but note that ten of these candidates have distances $8 \leq d < 10$~Mpc) and eighteen have distances $>45$~Mpc.

\begin{figure}
\includegraphics[scale=0.5]{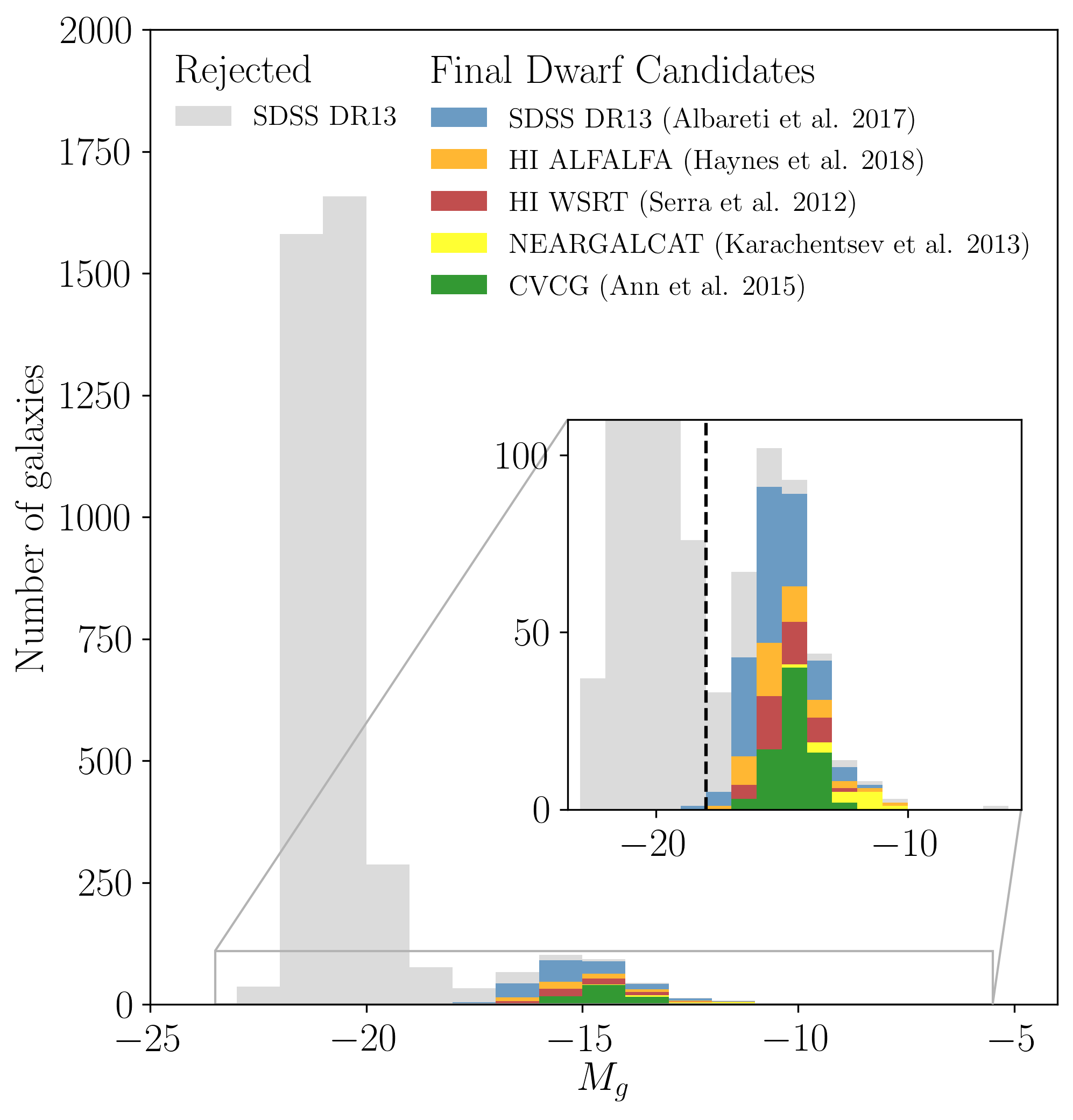}
\caption{Absolute \textit{g}-band magnitudes of the galaxies in our sample using distances determined from $z_{spec}$ measurements from SDSS DR13 with $S/N_{median} \geq 5$ (blue --- galaxies in our final dwarf catalogue; gray --- sources we rejected), a $z_{spec}$ measurement from the Catalogue of Visually Classified Galaxies \citep{Ann15}, a distance measurement in the Updated Nearby Galaxy Catalogue (NEARGALCAT; \citealt{Karachentsev13}), or HI narrowband imaging from the ATLAS$^{3D}$ survey \citep{Serra12} or ALFALFA \citep{Haynes18}. }
\label{figure:SDSScompare}
\end{figure}

\subsubsection{Absolute magnitudes of the dwarf candidates}
As a first test of the dwarf nature of our candidates, we used the distances calculated above to determine the absolute  {\textit{g}-band} magnitudes of the galaxies. In addition to our final sample of dwarfs, we also determined the absolute magnitudes of the 3862 rejected candidates with SDSS $z_{spec}$ measurements  in order to quantify the number of dwarf candidates that we missed. The $M_g$ values, seen in Figure~\ref{figure:SDSScompare}, show a bimodal distribution with a peak at $M_g \sim -21$ for the massive background galaxies and a smaller peak at $M_g \sim -15$ which is composed of dwarfs. All of our dwarf candidates lie between $-18.1 < M_g < -7.6$, and only one of the candidates is brighter than $M_g = -18$, suggesting a $<1$\% (1/292) contamination rate. Rather, the data suggests that we may have been too conservative, as we rejected 37 galaxies with $M_g > -16$. A subsequent visual inspection of these objects revealed, however, that many of these galaxies have extended nuclei or hints of spiral structure --- two features that we explicitly chose to reject from our sample. 

\begin{figure*}
\includegraphics[scale=0.33]{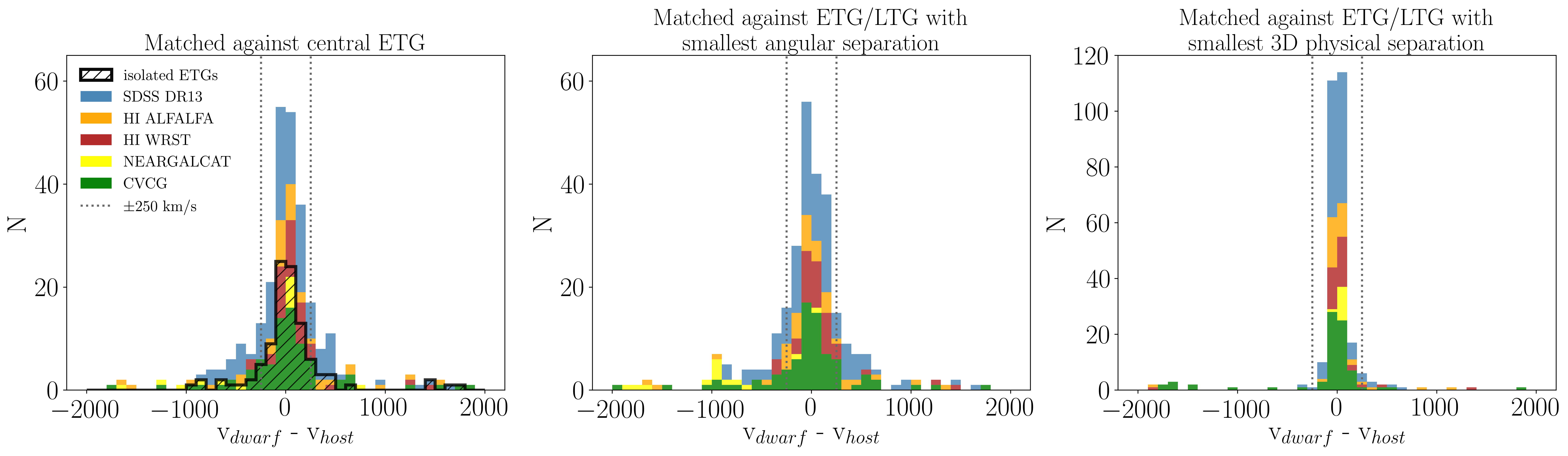}
\caption{The relative velocities of our dwarf candidates compared against the assumed host galaxy. The dwarfs were associated with a host galaxy using three different tests: (1) assuming the central ETG is the host, (2) matching dwarfs with the nearest on-sky neighbour (using only the RA and Dec) from the ATLAS$^{3D}$ catalogues, and (3) matching the dwarfs to the nearest neighbour in three-dimensional space using RA, Dec, and distance.  }
\label{figure:SDSScompare2}
\end{figure*}

\subsubsection{Association with the ATLAS$^{3D}$ massive galaxies}
\label{section:relv}
This sub-sample can also be used to test our assumption that the dwarf candidates form satellite populations around the more massive galaxies in the images, rather than field galaxies.  The distances indicators from Section~\ref{bfcontamination} were transformed into velocities where necessary, and compared against the $v_{helio}$ measurements for the early and late type galaxies in the ATLAS$^{3D}$ parent catalogue \citep{Cappellari11}; no corrections were applied to account for the peculiar velocities of any of the galaxies.

Operating under the assumption that our dwarf candidates are satellites of the massive galaxies in the ATLAS$^{3D}$ sample, each candidate is matched to a potential host galaxy using three different matching routines: (1) assuming all galaxies in each field are satellites of the central ETG --- this is likely to be a poor assumption, particularly in the fields with multiple ETGs/LTGs from the ATLAS$^{3D}$ parent catalogue, (2) matching the dwarfs to the massive galaxy with the smallest angular (on-sky) separation, and (3) matching dwarfs to the massive galaxy with the smallest physical 3D separation. In the latter two tests, it should be noted that the host galaxy can be a LTG, and the host does not have to fall within the MATLAS images. The third test is the most accurate, but the first two are included to test the accuracy of associating dwarfs to a potential host galaxy when the distance is unknown. The relative velocities of the dwarfs compared to their presumed hosts in each test are shown in Figure~\ref{figure:SDSScompare2}. If we assume that galaxies with $|\Delta v| < 250$~km/s, a typical velocity dispersion observed in Hickson compact groups \citep{Hickson97}, are satellites of the matched host galaxy, 88\% of the dwarfs in the third test are indeed associated with a nearby massive galaxy. However, the $|\Delta v|$ cut should not be a strict cutoff; galaxy groups with larger dispersions have been observed, the ATLAS$^{3D}$ galaxies span a range of masses and local densities that will affect the velocity dispersion of a satellite population, and the velocity estimates have associated errors.  If we instead adopt $|\Delta v| < 500$~km/s, 91\% of the dwarf candidates appear to be satellites of nearby massive galaxies.

The dwarfs that fall outside of these limits are more likely to be background dwarfs than foreground galaxies. Counting the dwarf candidates with $v_{dwarf} - v_{host} < -250$~km/s and $v_{dwarf} - v_{host} > 250$~km/s, the background to foreground contaminant ratio is roughly 3:2. This ratio is similar if counting the number of dwarfs outside $|\Delta v| < 500$~km/s, becoming 1:1. This finding is not unexpected, as we probe a larger volume of space at larger distances and should be able to detect dwarf galaxies at distances well beyond the artificial truncation at $d\sim45$~Mpc of the ATLAS$^{3D}$ catalogue.

It is immediately obvious from Figure~\ref{figure:SDSScompare2} that the first two satellite-host matching routines are not particularly accurate. Using $|\Delta v| < 250$~km/s, only 64\% of the dwarfs in the first test (70\% if we restrict the sample to dwarfs in fields with a single massive galaxy) would be identified as satellites. Only 63\% of the dwarfs are identified as satellites in the second test. Adopting the larger cutoff at $|\Delta v| < 500$~km/s, this fraction increases to 77~--~82\% in all three cases.

In future work, we will use surface brightness fluctuations (SBF) to estimate the distances to a large fraction of the dwarf candidates, which will allow us to more accurately pair satellites and host galaxies. Simulations by \citet{Mieske03} suggest that at a distance of 20~Mpc, one can use the SBF method to reliably determine distances for dEs as faint as $M_{V,lim}=-10.8$ [$-12.75$] in images with a one hour integration time and 0.5$\arcsec$ [1.0$\arcsec$] seeing; this limiting magnitude becomes  $M_{V,lim} =~-14.9$ [$-16.75$] at $d \sim 50$~Mpc.

\section{Results}
\label{results}
\subsection{The number of dwarfs as a function of host properties}
\label{section:rho10}
\begin{figure*}
\centering
\includegraphics[scale = 0.3]{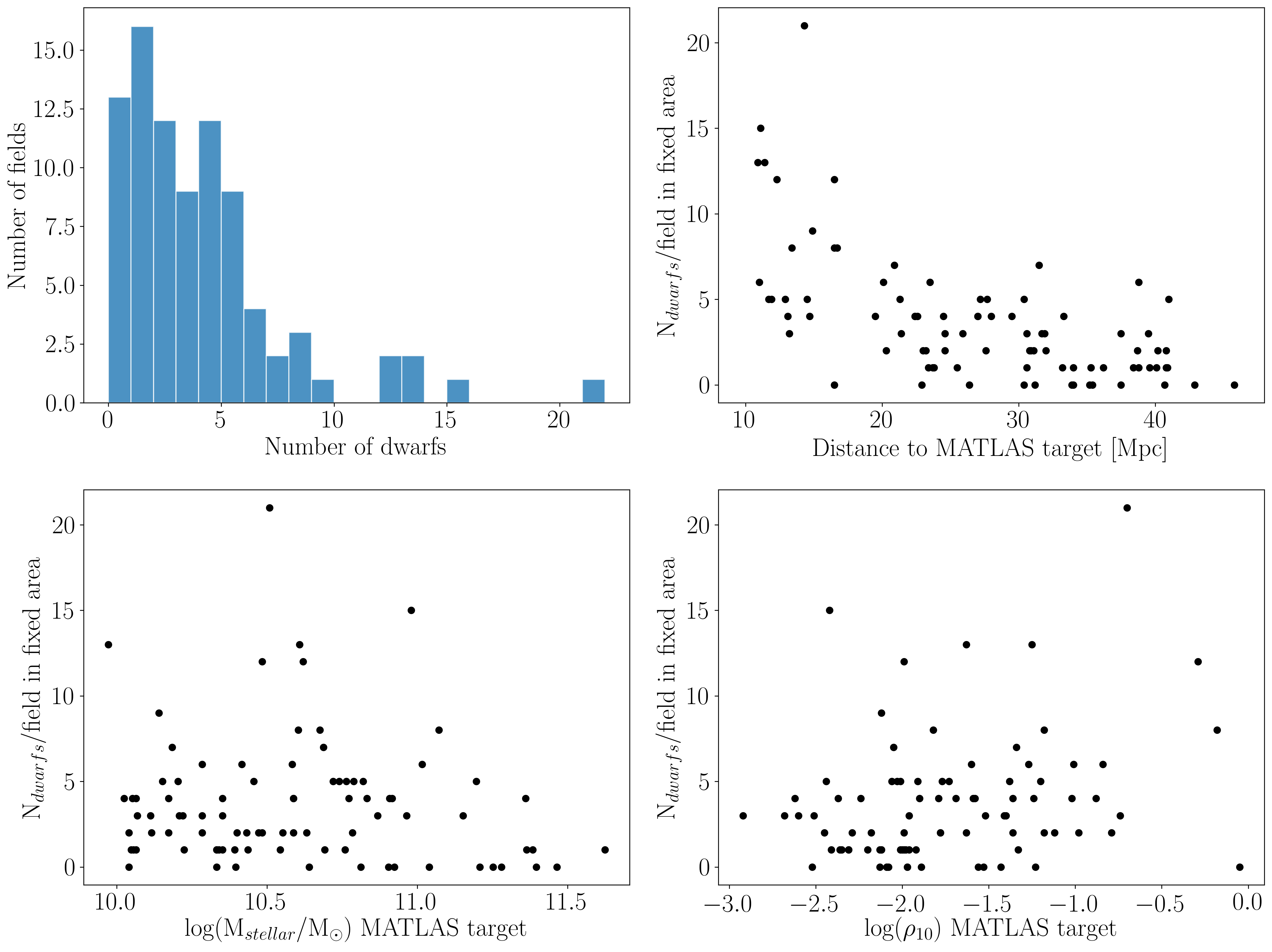}
\caption{\textit{Top Left:} The number of dwarfs per field, if we restrict the sample to the 87 fields containing an isolated ETG and count the number of dwarfs only within a fixed region ($175\times 175$~kpc) centred on the MATLAS target. \textit{Top Right:} The same sub-sample as left, plotted versus the distance to the MATLAS target.  \textit{Bottom Left:} The number of dwarfs versus the stellar mass of the MATLAS target. \textit{Bottom Right:} The number of dwarfs versus the local density, $\rho_{10}$, of the MATLAS target. The masses and $\rho_{10}$ values were taken from \citet{Cappellari11,Cappellari11b}.  }
\label{figure:dwarfsperfield}
\end{figure*}

The total number of dwarfs we identified within each MATLAS image is plotted in the rightmost histogram of Figure~\ref{figure:dwarf_separation}. It should be noted, however, that this number is not expected to trace any natural variation in the number of dwarf satellites between systems, which recent observations suggest may depend on the morphology and/or the mass of the host galaxy \citep{Wang12,Nierenberg12,Nickerson13,Ruiz15,Ann15}. Although it counts the number of dwarfs within a fixed angular region of the sky, it does not take into consideration the different physical scales probed in each image, the mass of the host galaxy, or the number of host ETGs/LTGs within the fields. A better tracer would be the number of dwarfs found in a fixed physical area of the sky centred on the ETGs with no other massive galaxies in close projection. Eighty-seven ETGs satisfy this requirement, and the search area was determined by the physical size of the nearest field (a square $\sim175$~kpc on a side). In total, there are 729 dwarf candidates within these regions, with individual fields containing 0~--~23 dwarfs (the median is five candidates/field). The distribution is shown in Figure~\ref{figure:dwarfsperfield}.  

The variation in $N_{dwarfs}$ as a function of the distance to the MATLAS target, the stellar mass of the host, and the local density are also included in Figure~\ref{figure:dwarfsperfield}. The local density $\rho_{10} = 3 N_{10}/(4\pi r_{10}^3)$ where $N_{10}=10$ is the number of nearest neighbours used in the test and $r_{10}$ is the distance to the 10$^{th}$, using the massive galaxies in the ATLAS$^{3D}$ parent catalogue \citep{Cappellari11b}. As expected, the number of dwarfs falls as a function of distance, but we do not observe any correlation between $N_{dwarfs}$ and the stellar mass or the local density of the ETG. This is in direct conflict with simulations and prior observations; simulations suggest that we should find $\sim 5\times$ more dwarfs around the most massive ETGs in the sample than the least massive galaxies (e.g., \citealt{Javanmardi19}, and a study of dwarfs in the COSMOS fields by \citet{Nierenberg12} found that more massive ETGs (in a sample spanning masses $10.5 \lesssim \log(M_{*}/M_\odot) \lesssim 11.5$) host significantly more satellites than less massive ETGs. The discrepancies are likely the result of the limited region in which $N_{dwarfs}$ was counted and competing effects (e.g., mass, density, morphology) which have not been fully disentangled.

In principle, we can also explore the variation in $N_{dwarfs}$ as a function of the host morphology, as the MATLAS images do contain a number of LTGs. However, this depends on how accurately one can disentangle superimposed satellite populations and correctly associate the satellite and host galaxies. In the fields containing a single ETG, we are necessarily restricted to hosts with morphological types E, S0, or S0a. Interestingly, we do observe systematic differences in the number of dwarfs around these hosts. Around the massive Es, the median number of dwarfs is 9.5, while this drops to 6 around the S0 galaxies and 5 around the S0a galaxies. This trend agrees with the findings of \citet{Ruiz15}, although they counted $N_{dwarfs}$ differently by not restricting the sample to dwarfs within a fixed physical around the host and counting the satellites down to a mass ratio of 1:100 of the host.

\begin{figure*}
\centering
\includegraphics[width=0.95\textwidth]{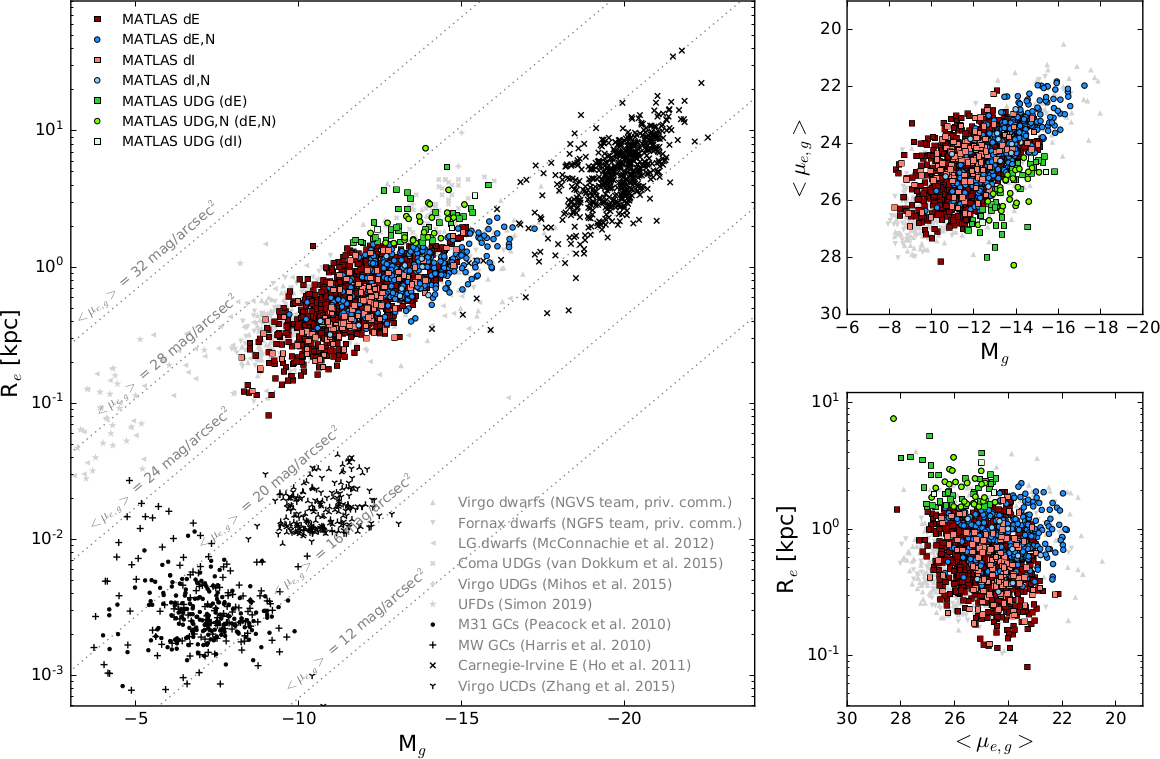}
\caption{Testing three scaling relations for our dwarf candidates: $M_g$ vs.\ $R_{e}$ (\textit{left}), the average surface brightness within $R_{e}$ ($\langle \mu_{e,g} \rangle$) versus $M_g$ (\textit{right, top}), and  $\langle \mu_{e,g} \rangle$ versus $R_{e}$ (\textit{right, bottom}). The plotted quantities are based on parameters returned by the {\sc{Galfit}} modelling, and we have assumed the dwarf candidates are at the same distance as the targeted ETG in the field (the first matching routine in Section~\ref{section:relv}). The nuclei have been masked, and only the properties of the diffuse component of the galaxy are used. Our candidates occupy the same space as dwarf candidates identified in the NGVS (grey triangles) and NGFS (inverted grey triangles; see \citealt{Eigenthaler18} for a discussion) images, ultra diffuse galaxies in the Coma cluster (grey pentagons) and Virgo cluster (grey octagons), as well as Local Group dwarfs (rotated grey triangles). For comparison, we have also included ultra compact dwarfs (black `tri\_down' symbols), ultra faint dwarfs (grey stars), globular clusters around the Milky Way (black pluses) and Andromeda (black circles), and massive ellipticals from the Carnegie-Irvine catalogue (black x's). }
\label{figure:fundamentalplane}
\end{figure*}

\subsection{Scaling Relations}
The basic properties of our sample can be viewed in Figure~\ref{figure:fundamentalplane}, where we confirm that our candidates follow the standard scaling relations between $M_g$, $R_{e}$, and $\langle \mu_{e,g} \rangle$. In order to convert $R_e$ into physical units and calculate the absolute magnitudes, we have assumed that the dwarf candidates are at the same distance as the targeted ETG in each field (the first matching routine discussed in Section~\ref{section:relv}). All three parameters used in these plots are from the {\sc{Galfit}} modelling and the candidates without good fits --- typically the smallest and faintest candidates --- have been removed from the sample. Here we present the results for the remaining 1474 galaxies: 1002 dEs, 329 dE,Ns, 133 dIrrs, and 10 dIrr,Ns. It should be noted that any nuclei were masked prior to running {\sc{Galfit}}, and the plotted parameters correspond to only the diffuse component of the galaxies.

\begin{figure}
\centering
\includegraphics[width=0.48\textwidth]{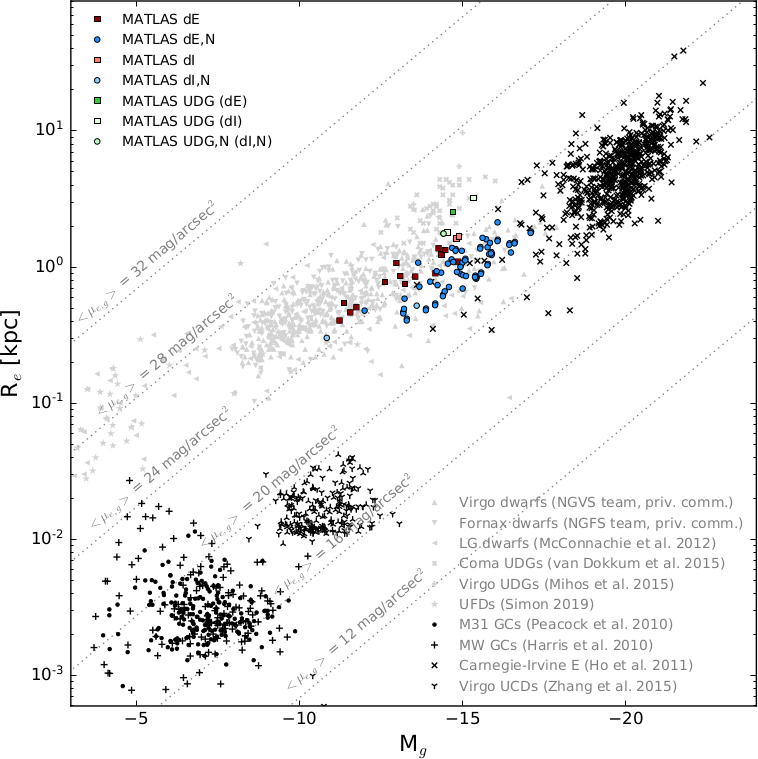}
\caption{Same as Figure~\ref{figure:fundamentalplane}, but showing only the subsample of dwarf candidates with a distance measurement. The parameters were recomputed using this distance rather than the distance to the central ETG in the field.  }
\label{figure:fundamentalplane2}
\end{figure}

For comparison, several other galaxy catalogues are also plotted in Figure~\ref{figure:fundamentalplane}. These include dwarfs from the Local Group \citep{McConnachie12}, dwarfs identified in the  NGVS (NGVS team, private communication) and the NGFS (NGFS team, private communication) images, massive Es \citep{Ho11}, UCDs \citep{Zhang15,Liu15}, ultra diffuse galaxies \citep{vanDokkum15,Mihos15}, UFDs \citep{Simon19}, as well as globular clusters \citep{Harris10,Peacock10}. The NGVS and NGFS are complementary, with image quality and depth comparable to the MATLAS images. We therefore expect to find similar dwarf populations in all three surveys --- barring potential differences due to the environments probed --- and in fact the dwarf populations from all three overlap almost perfectly (the NGVS and NGFS samples are seen more clearly in Figure~\ref{figure:fundamentalplane2}, see discussion below). It should be emphasized once more that we do not expect to find UCDs in our sample, not because they are not present in the images, but rather because they are not expected to survive the area cut we applied during the candidate selection. Properties of the UCD and globular cluster populations present in the MATLAS images will be presented in a separate paper (Lim et al., in preparation). These scaling relations serve as an additional sanity check on our sample, and suggest once more that we have constructed a robust catalogue of dwarf candidates.

As stated above, to create Figure~\ref{figure:fundamentalplane} we had to assume a distance to the dwarfs.  To test the impact of this assumption, we recalculated the absolute magnitudes and effective radii using the distance indicators described in Section~\ref{bfcontamination}, and highlight this sub-sample in Figure~\ref{figure:fundamentalplane2}. All of the candidates again fall within the space inhabited by other dwarf samples. Interestingly, we see a separation between the nucleated and non-nucleated dwarfs in the plot, such that at a fixed magnitude the nucleated dwarfs have smaller effective radii. There is also tentative evidence from the NGVS dwarfs for such a trend \citet{SanchezJanssen18}, although the authors note that the finding is not statistically significant given their sample size. This will be re-investigated once we have a larger number of dwarfs with confirmed distances, to test if this effect is real, but obscured in Figure~\ref{figure:fundamentalplane} due to uncertainties in the distances, or the result of selection biases.

\subsubsection{Ultra Diffuse Galaxies}
\label{section:udgs}
Our sample includes a number of extended, low surface brightness galaxies.  Examples of such objects have been known for decades \citep{Binggeli84,Impey88,Bothun91,Turner93,Dalcanton97,Caldwell06}, but it was not until they were defined as ultra diffuse galaxies (UDGs) in 2015 that they became the focus of renewed attention \citep{vanDokkum15}.  To identify the UDGs, we adopt the parameter cuts defined by \citet{vanDokkum15}: $R_e \gtrsim 1.5$~kpc and $\mu_{0,g} \gtrsim 24$~mag/arcsec$^2$. The surface brightnesses $\mu_{0,g}$, $\mu_{e,g}$ and $\langle \mu_{e,g} \rangle$ were calculated using the total magnitude and effective radius returned by {\sc{Galfit}} and assuming a Sersic profile with n=0.8, the median of our sample.

The juxtaposition of dwarf (low surface brightness) and non-dwarf characteristics (large sizes) leaves the nature of UDGs open to interpretation. It has been suggested that UDGs are dwarf galaxies with an unusually large extent \citep{Roman16}, which may have been generated by high spins \citep{Amorisco16}, central collisions between galaxies \citep{Baushev18}, strong gas outflows associated with star formation \citep{DiCintio17}, tidal stripping \citep{Carleton18}, or a tidal-dwarf formation scenario \citep{Duc14,Bennett18}. Others, however, support the view that UDGs are similar to massive galaxies (i.e., that they formed in massive halos typically populated by more massive galaxies)  but have unexpectedly low surface brightnesses \citep{vanDokkum16}. Observations to estimate the masses of UDGs have been ambiguous; colour-magnitude relations and gravitational lensing studies suggest that UDGs are, on average, low mass systems \citep{Beasley16,Sifon18}, but a kinematic study of one UDG reveals that it is a massive galaxy \citep{vanDokkum16}, and estimates based on counts of globular cluster populations indicate that some are massive \citep{vanDokkum16,vanDokkum17} while others are not (\citealt{Amorisco18}; see also \citealt{Lim18}). The position of the UDG candidates in our scaling relations plots suggests they are not a distinct class of objects, but rather an extension of the dwarf population.

Thus far, most UDGs have been identified in clusters, but a growing number of them have been identified in low density environments (e.g., \citealt{Crnojevic16,Roman16,Leisman17,Papastergis17,Jiang18,Muller18c}). Characterizing this population  is important to better understand the formation and evolution of these galaxies. Properties of the MATLAS UDGs will be explored in detail in an upcoming paper (Marleau et al.\, in preparation).

\subsection{Dwarf colours}
The detection of the dwarf candidates was based solely on the g-band imaging, but MATLAS obtained multi-band imaging of most fields, allowing us to extract $(g-r)$ and  $(g-i)$ colours for a fraction of the candidates. Fluxes were extracted within $R_e$, taken from the {\sc{galfit}} modelling of the galaxies on the g-band imaging, using the python \textit{photutils} package \citep{Bradley19}. The galaxies without {\sc{galfit}} modelling have been dropped from this analysis.  A 1.5\arcsec{} region around all nuclei was also masked, to prevent contamination of the host galaxy colours. There are 1270 galaxies with detections in both \textit{g} and \textit{r}, and 667 galaxies with detections in \textit{g} and \textit{i}.

The magnitudes were corrected for Galactic extinction and K-corrections were applied to remove any distance dependence. Extinction corrections were obtained from the IRSA database at the coordinates of each candidate, using the reddening values from \citet{Schlafly11} and assuming $R_V=3.1$.  The median \textit{g}, \textit{r}, and \textit{i}-bands extinction corrections are 0.09, 0.06, and 0.04, respectively. The K-corrections were applied following the method outlined in \citet{Chilingarian10}; as expected, given the distances of the MATLAS targets, the correction factors are negligible, with median values 0.003, 0.008, and 0.002 magnitudes in the \textit{g}, \textit{r}, and \textit{i}-bands, respectively. Applying these corrections, we measure median colours $(g-r)_0 = 0.49$ and $(g-i)_0 = 0.78$ for the full sample.

\begin{figure*}
\centering
\includegraphics[width = \textwidth]{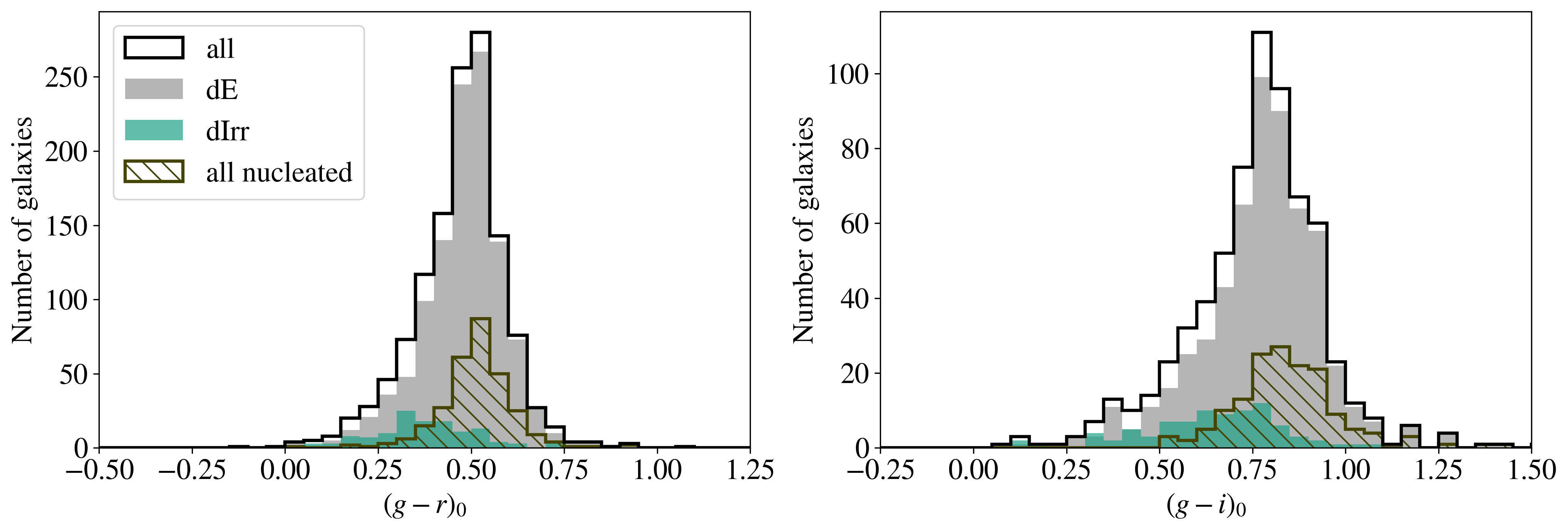}
\caption{The extinction corrected $(g-r)_0$ (\textit{left}) and $(g-i)_0$ (\textit{right}) colours of our dwarf candidates, separated by morphological type and nucleated status. The colours are measured within a fixed aperture with radius $1R_e$. }
\label{figure:color1}
\end{figure*}

The corrected colours, subdivided into the different morphological classes (dE, dIrr, and nucleated candidates) can be seen in Figure~\ref{figure:color1}.  As expected, the dIrr are bluer on average than the dE candidates. Interestingly, however, we do find dE candidates that are as blue as the dIrr galaxies. The median extinction-corrected $(g-r)_0$ colours for the dE and dIrr galaxies are 0.49 and 0.37, respectively, while the median $(g-i)_0$ colours of the two morphology classes are 0.79 and 0.65. Additionally, the nucleated dwarfs are redder than the general population of dwarfs. The median colours of the nucleated and non-nucleated population (not explicitly shown in Figure~\ref{figure:color1}) are 0.52 and 0.47 in $(g-r)_0$, and 0.83 and 0.76 in $(g-i)_0$. This dichotomy is likely explained by the observation that nuclei are more commonly found in brighter dwarf galaxies (see Figures~\ref{figure:fundamentalplane} and \ref{figure:nucfraction} and the discussion in Section~\ref{section:nucfraction}) and that brighter galaxies tend to have redder colours (e.g., \citealt{Roediger17}).

These results appear to be in good agreement with $(g-r)_0$ colours reported in the literature. For example, the median global color, measured in SDSS filters, of dwarfs in the Centaurus~A group is $(g'-r')_0 = 0.463$ \citep{Muller17a}, $(g'-r')_0 = 0.472$ in the M101 group \citep{Muller17}, and $(g'-r')_0 = 0.491$ in the Leo-I group \citep{Muller18c}. However, care needs to be taken when comparing these values. We measure colours within $1R_e$, rather than the global colours. Although dwarfs are expected to have shallow colour gradients --- for example, \citealt{Urich17} measured a median colour gradient in $g-r$ of $-0.006$~mag/$R_e$ for a sample of dwarfs in the Virgo Cluster --- it is unclear at present if, and by how much, measurements of the global colours would change the median values of our sample. Additionally, the MegaCam and SDSS filters are not identical, leading to additional uncertainties.

Somewhat surprisingly, our $(g-i)_0$ colours also appear to be in fairly good agreement with the colours of dwarfs found in the Fornax Cluster by NGFS. \citet{Eigenthaler18} measure a median value $(g-i)_0 \sim 0.8$, compared against 0.78 for our sample. However,  Eigenthaler et al.\ note themselves that a comparison sample taken from NGVS (which also utilized MegaCam) is systematically redder than the NGFS sample, which may be the result of uncertainties in the transformations between the photometric filters of the two instruments. If so, and the NGFS dwarfs are `corrected' to redder values, the colour difference between our sample of dwarfs in low density environments and those in clusters would increase, with the MATLAS dwarfs bluer, as expected.

\subsection{Stellar mass estimates}
Using the colours above and assuming a distance to the dwarf candidates, it is possible to estimate their stellar masses.  \citet{Taylor11} derived an empirical relation between the $(g-i)$ colour, restframe \textit{i}-band absolute magnitude, and stellar mass by modelling the spectral energy distributions (SEDs) of galaxies in the Galaxy And Mass Assembly (GAMA) survey. The final stellar mass estimates are based on the stellar population synthesis (SPS) models of \citet{Bruzual03}, and assume a \citet{Chabrier03} initial mass function (IMF) and \citet{Calzetti00} dust law. It should be noted that while the GAMA sample is dominated by massive galaxies, it does include dwarfs with $M_{stellar} < 10^9 M_\odot$, as well. For this calculation, we measured the \textit{i}-band flux within $3R_e$ as a proxy for the total flux of the galaxy, and assumed the dwarf candidates are at the same distance as the central ETG in the image. Furthermore, we dropped galaxies with bright neighbours within $3R_e$ that would affect the measured flux, which reduced the sample to 658 galaxies.

\begin{figure}
\captionsetup{width=0.9\textwidth}
\centering
\includegraphics[width=0.48\textwidth]{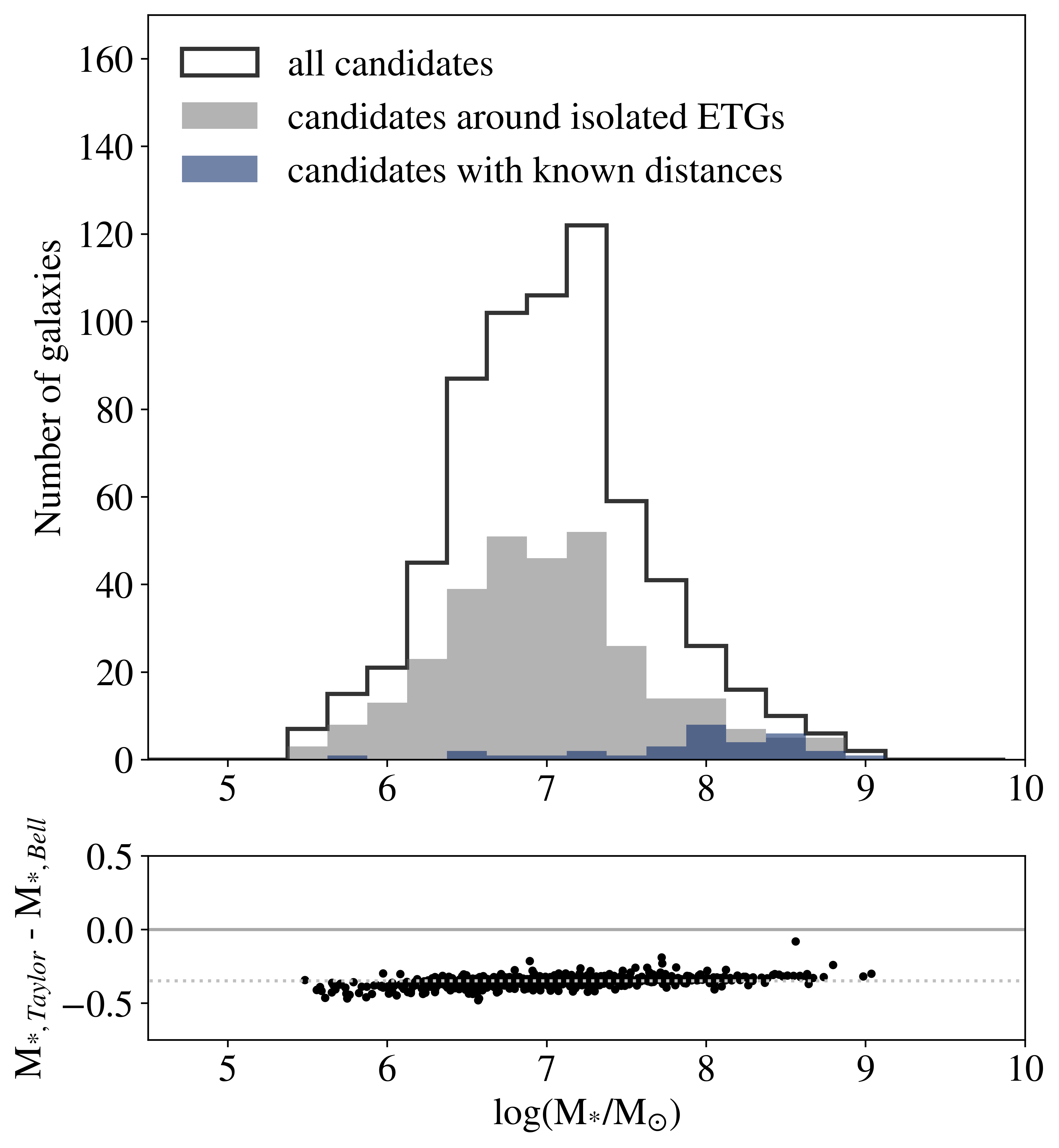}
\caption{\textit{Top:} The distribution of stellar masses for the 658 galaxies with $(g-i)$ colour estimates (shown in Figure~\ref{figure:color1}) and clean photometry, based on the \citet{Taylor11} colour-mass relation. See text for details. Dwarfs in fields containing a single massive galaxy, whose distances are expected to be more robust, are shown in gray. Mass estimates were also re-computed for the dwarfs with known distances using the measured values (blue). \textit{Bottom:} A comparison of the stellar masses returned by the  \citet{Taylor11} and \citet{Bell03} colour-mass relations as a function of the Taylor et al.\ mass.  }
\label{figure:massestimates}
\end{figure}

The resulting stellar masses are shown in Figure~\ref{figure:massestimates}. In log space, the masses range from $5.5 < \log(M*/M_\odot) \leq 9.0$, with a median value of 7.0$^{+0.1}_{-0.2}$, suggesting once more that our low surface brightness sample is dominated by low mass galaxies.  The error was calculated assuming an uncertainty of $\pm3.5$~Mpc in the distance to each dwarf candidate (estimated from the dispersion in the relative velocities shown in Figure~\ref{figure:SDSScompare2}, left).

The uncertainty in the distances of the dwarfs is expected to dominate the errors in the mass estimates. To check the impact of the distance, we highlight two sub-populations in Figure~\ref{figure:massestimates}: the 306 dwarf candidates in fields containing an isolated ETG and the 32 candidates with prior distance measurements. As discussed previously, the association between dwarfs and the central ETG is slightly more robust in the fields containing a single massive galaxy and therefore the assumed distance is expected to be a better estimate; however, this does not significantly alter the distribution of stellar masses and the median value is unchanged at 7.0. For the candidates with known distances, the stellar masses were re-estimated using the measured values. The brighter dwarfs are more likely to have a prior distance estimate, and it is therefore unsurprising that the distribution of stellar masses for this sub-sample is slightly skewed towards higher masses, with  values ranging from $5.8 \leq \log(M_*/M_\odot) \leq 9.0$ and a median stellar mass in log space of 7.9. 

In the literature, stellar masses are also commonly estimated using the colour-mass relations derived by \citet{Bell03}. They present a series of colour-mass relations for colours and absolute magnitudes in various optical bandpasses, based on SED modelling of a galaxies in the SDSS Early Data Release. The relations are similar to the Taylor colour-mass relation described above, but there are two key differences: Bell et al.\ adopt a `diet' \citet{Salpeter55} IMF by scaling the mass-to-light ratios in the standard Salpeter IMF by a factor of 0.7,  and they apply an additional `evolutionary correction' by running the SPS models forward to $z=0$. We directly compare the results of the two stellar mass estimates in the lower panel of Figure~\ref{figure:massestimates}; using the same $(g-i)$ colours and $M_i$ values, the Bell relation returns mass estimates that are typically 0.35~dex more massive than the Taylor relation. As a result, five of our dwarf candidates have stellar mass estimates above $10^9 M_\odot$, with the most massive systems at $\log(M_*/M_\odot) = 9.3$, which is still below the mass of the LMC ($\sim3\times10^9$; \citealt{Marel04}). This difference can be attributed to the different IMFs; SPS modelling indicates that a standard Salpeter IMF will return a stellar mass $\sim$0.24~dex more massive than a Chabrier IMF (e.g., \citealt{Mitchell13}), which increases to $\sim$0.34~dex for the diet Salpeter IMF.

\subsection{The dwarf morphology-density relation}
\label{section:morph-density}
\begin{figure}
\centering
\includegraphics[scale = 0.45]{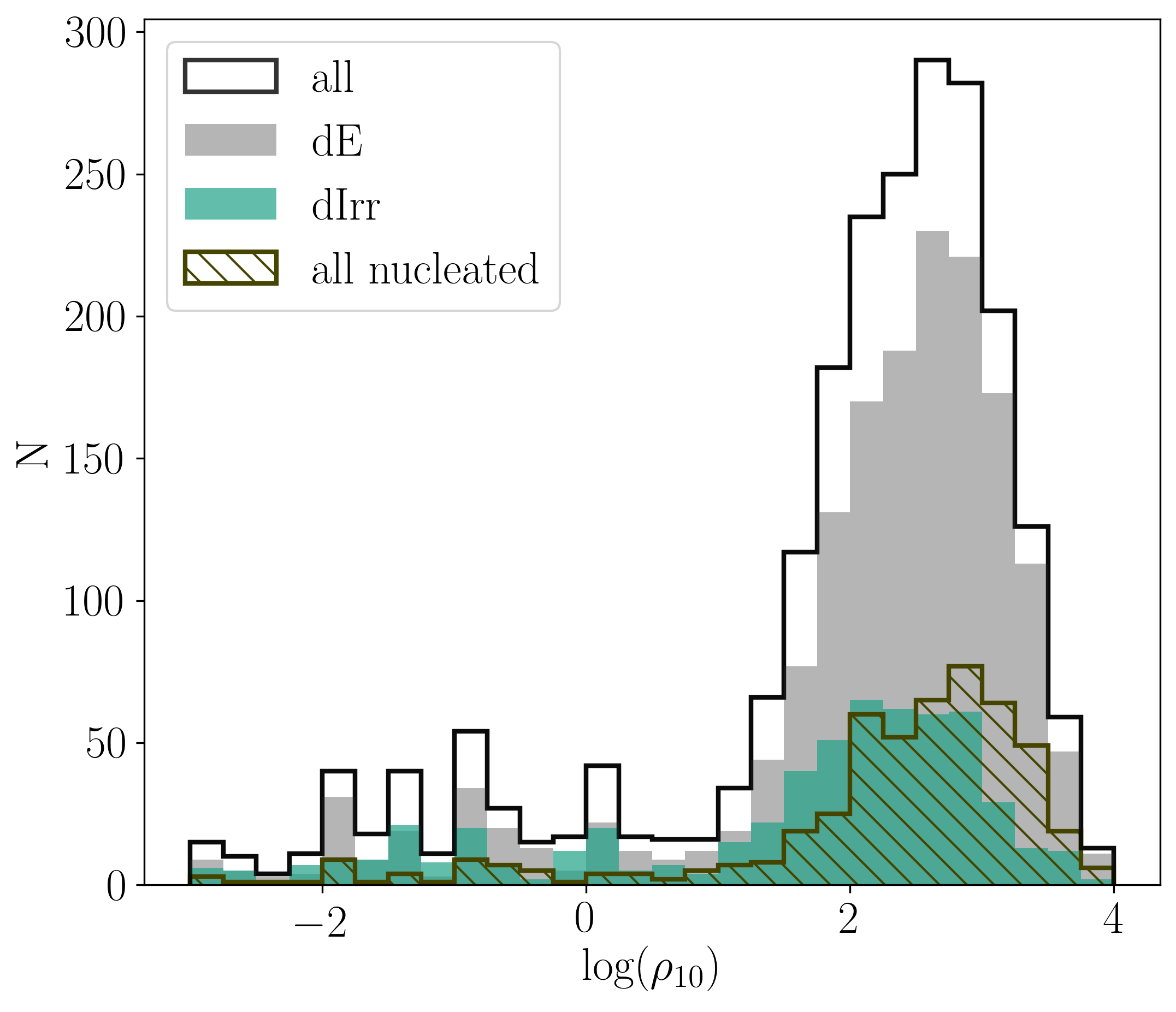}
\caption{The local densities of the dwarf candidates in our sample. Here we make the assumption that the dwarfs in a given field are at the same distance as the central ETG, thus the $r_{10}$ separation is likely underestimated and the final $\rho_{10}$ values are overestimates. However, the distribution of nucleated dwarfs is shifted towards higher density regions compared to the overall dwarf population, while the distribution of dIrrs is shifted towards lower density regions.  }
\label{figure:localdensity}
\end{figure}

Observations of dwarfs in nearby groups and clusters show that dwarf galaxies follow a morphology-density relation similar to massive galaxies, such that dEs are preferentially located in denser environments and dIrrs in lower density environments  \citep{Dressler80,Ferguson90,Grebel03,Skillman03,Cote09,Weisz11,Karachentsev15b}. However, it is unclear if a group (or more dense) environment is a necessary condition to establish the morphology-density relation, or if it also exists around isolated host galaxies. To test for the presence of a morphology-density relation in the MATLAS sample, we plot the distribution of the morphological classes as a function of the local density ($\rho_{10}$) of each candidate. The density was calculated following the same definition discussed in Section~\ref{section:rho10}, but using all entries in the ATLAS$^{3D}$ parent catalogue as well as our dwarf sample. The numerical values should be treated with caution --- we have assumed the dwarfs are at the distance of the central ETG, thus $r_{10}$ is almost certainly underestimated, but it may be overestimated near the boundaries of the ATLAS$^{3D}$ survey where the catalogues artificially truncate  --- but the separation of high and low density regions (larger and lower $\rho_{10}$ values, respectively), should still have some meaning. The distribution can be seen in Figure~\ref{figure:localdensity}.

The dIrr candidates can be found throughout the $\rho_{10}$ distribution, but are shifted towards lower density environments. The median density of the dIrr galaxies is $\rho_{10} = 2.1$, versus 2.4 for the full sample. This trend is even clearer if one considers the relative fraction of dIrr candidates in three regions: $\rho_{10} < 1.0$, $1.0 \leq \rho_{10} < 3.0$, and $\rho_{10} \geq 3.0$. In the lowest density range, the irregular dwarfs account for  41\% of all dwarfs in the region. In the next bin, $1.0 \leq \rho_{10} < 3.0$, the irregular fraction decreases to 20\%, and this falls to 14\% when $\rho_{10} \geq 3.0$. Again, the $\rho_{10}$ values should be treated with caution, but this trend supports previous observations that dIrr galaxies are more commonly found in less dense regions. 

For the subsample of dwarf candidates with known distances, we can also directly calculate the separation between the dwarfs and their hosts to check for any systematic variations in the distributions of the morphological subtypes. Assuming the host galaxy is the ETG/LTG with the smallest physical separation, and rejecting the dwarf candidates at $d>5$~Mpc to remove the outliers, we do observe that the dE candidates are located closer to the host, on average, than the dIrr candidates. The median separation of the dEs is 0.7~Mpc, while the median separation of the dIrrs is 1.0~Mpc. \citet{Skillman03} and \citet{Cote09} have performed a similar calculation for dwarfs in the Sculptor Group and in Centaurus A, respectively; while our value for the median distance of the dIrr candidates is in rough agreement with the other two studies (0.95~Mpc and 0.85~Mpc, respectively), the median separation of our dE candidates is almost three times larger than the values measured in the other two systems (0.22~Mpc and 0.23~Mpc, respectively). Although it is possible that we have missed a population of small dEs near the massive hosts, which would lower the median separation between the two, we would have had to miss a substantial population to reconcile our value with the previous studies. Rather, it seems likely that the morphology-density relation is not as clearly established in low density environments as in groups.  

\subsection{Nucleated Dwarfs}
\label{section:nucfraction}
A large number of dwarf galaxies have been observed with a compact central nucleus. The nature of these structures is ambiguous; many are consistent with massive, nuclear star clusters (e.g., \citealt{Oh2000,Ordenes18b,SanchezJanssen18}) although a growing number of central massive black holes have been found in dwarf galaxies (e.g., \citealt{Mezcua16,Mezcua18,Marleau17}) and it is likely that both compact objects populate the central region of the galaxies (e.g., \citealt{Georgiev16}). Many of the properties of the nuclei are found to scale with properties of the host galaxy, suggesting that the evolution of the two structures are linked. For example, the fraction of nucleated dwarf galaxies increases in brighter dwarfs, reaching 90\%~--~100\% for dwarfs in the Fornax and Virgo Clusters \citep{Munoz15,Ordenes18a,Ordenes18b,SanchezJanssen18}.

\begin{figure}
\centering
\includegraphics[scale = 0.45]{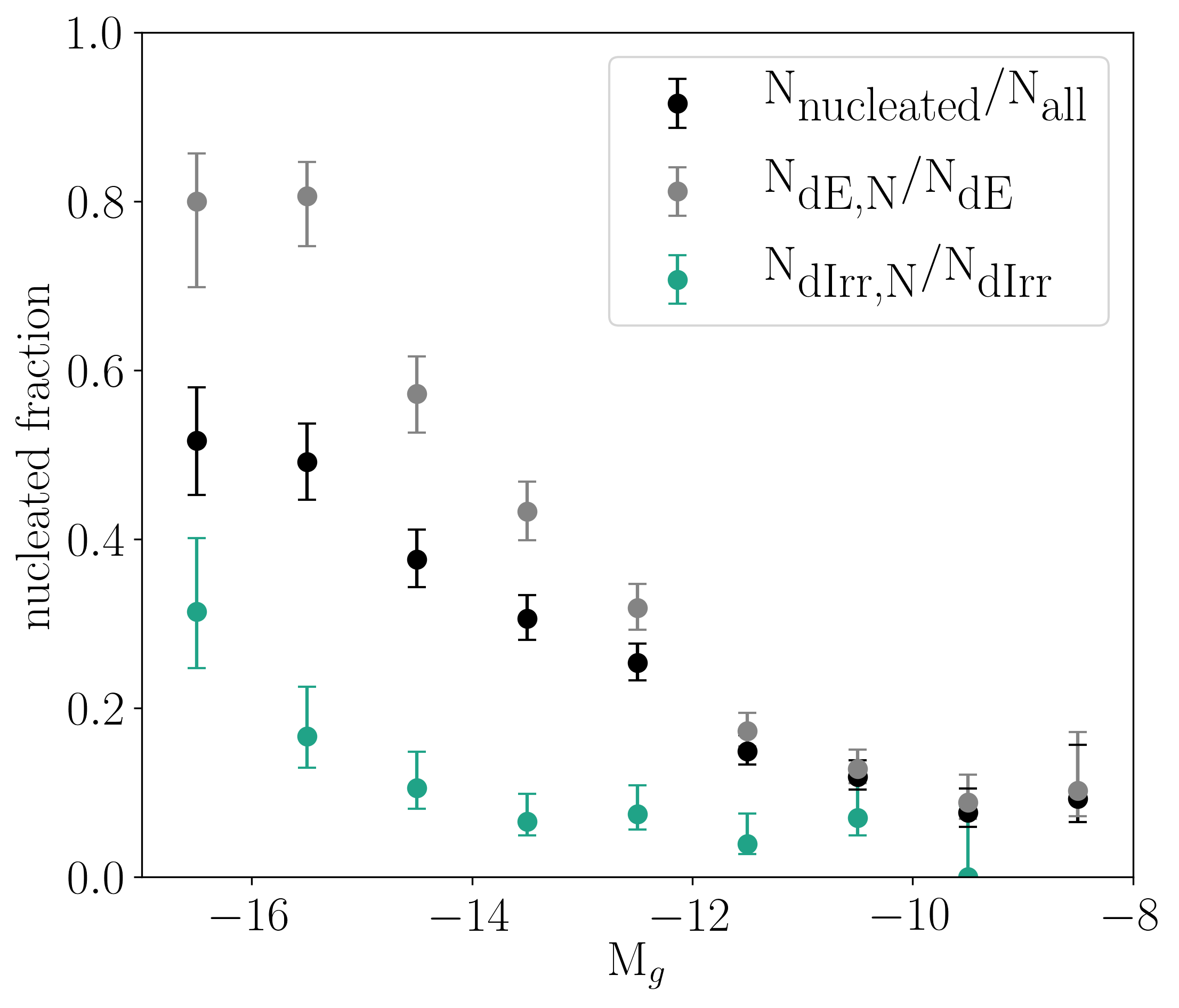}
\caption{The fraction of nucleated dwarfs as a function of absolute magnitude $M_g$ of the dwarf host for all candidates (black), within the dE subsample (gray) and within the dIrr population (turquoise). All three populations show a similar trend, such that the brighter dwarf galaxies are more likely to be nucleated, in agreement with the literature (e.g., \citealt{Munoz15,Ordenes18a,Ordenes18b,SanchezJanssen18}). The error bars correspond to 1$\sigma$ binomial confidence intervals.}  
\label{figure:nucfraction}
\end{figure}

This relation can be tested in low density environments using the MATLAS dwarfs. The fraction of nucleated dwarf candidates is plotted as a function of $M_g$ in Figure~\ref{figure:nucfraction} for the full dwarf sample (black points), as well as the nucleated fraction within the dE (gray points) and dIrr (turquoise points) subsamples. The error bars represent 1$\sigma$ binomial confidence intervals of the measured quantities \citep{Cameron11}. In all three samples, the nucleated fraction increases for the brighter galaxies, reaching 51.6\%, 80.6\%, and 33.3\% at $M_g \sim - 17$, for the full dwarf sample, the dE sub-sample, and the dIrr sub-sample, respectively. These percentages should be taken as lower limits, however, as our sample is likely missing some fraction of nucleated dwarfs at the bright end due to the confusion between bright nucleated dwarfs and background S0 galaxies. Indeed, the nucleated fraction we calculate for the dE sub-sample is slightly smaller, although consistent within the errors, of the $\gtrsim 90$\% reported for dwarfs in the Virgo and Fornax clusters \citep{Munoz15,Ordenes18a,Ordenes18b,SanchezJanssen18}.

Interestingly, we observe a slight flattening in the nucleated fraction at the $M_g \leq -16$ for the full sample and the dE candidates. This is consistent with the turnover observed by \citet{SanchezJanssen18}, who observe a peak in the nucleated fraction for galaxies with stellar masses $\sim10^9 M_\odot$.

Recently, particular attention has been placed on the nucleated fraction of the faint dwarf candidates. Pinpointing a magnitude (or mass) limit below which nuclei are not observed may place constraints on the formation and evolution of these structures. Previous studies have placed this limit at $M_g \sim -10$ or $\log(M_{stellar}/M_\odot) \sim 6.4$ \citep{SanchezJanssen18,Eigenthaler18,Ordenes18b}. We have identified a small number of dwarfs (15) in fainter dE candidates, although many of these are just below a $M_g = -10$ cutoff. A more in depth analysis of the nuclei, based on {\sc{Galfit}} modelling of the nuclei and host, will be presented in a future paper.

Previous observations have also found that nucleated dwarfs are more commonly found in higher density environments (e.g., \citealt{Binggeli91,Oh2000}). Although the overall nucleated fraction of the MATLAS dwarfs agrees with those found by the NGVS and NGFS teams, we can probe how this number varies with the local densities of the dwarfs by applying a calculation similar to that used in Section~\ref{section:morph-density} to test the morphology-density relation. The peak of the nucleated sub-sample is shifted towards higher density environments compared to the full sample, with a median value $\rho_{10} =  2.6$. Dividing the distribution into the same three regions used previously, the nucleated candidates account for 16\%, 17\%, and 34\% of the total galaxies at $\rho_{10} < 1.0$, $1.0\leq \rho_{10} < 3.0$, and $\rho_{10} \geq 3.0$, respectively. Thus, our data supports the trend observed that nucleated dwarfs are more concentrated in higher density environments. Recent work by \citet{SanchezJanssen18}, however, has found that the environmental dependence is actually secondary to the mass dependence; these effects will be disentangled in the future paper on the nucleated dwarfs.

\section{Conclusions}
\label{conclusions}

We have presented a first look at a new sample of 2210 dwarf galaxies identified in the MATLAS deep imaging survey. Our dwarf candidates were identified through a direct visual inspection of the images and by visually cleaning an automated catalogue generated using SExtractor parameters. SExtractor has limitations for this type of project --- it was not designed specifically for the detection and extraction of low surface brightness features --- and some fraction of the dwarfs will be missed, but we estimate that approximately 90\% of the dwarf candidates were extracted based on the candidates that were identified in the full-visual catalogue. The automated catalogues returned many potential dwarfs that were not included in the fully visual catalogue, however, and we ultimately identified $\sim$1.5$\times$ more dwarfs by adopting the semi-automated dwarf selection. It should be noted, however, that our catalogue is still incomplete; we have likely missed a number of dwarfs with small angular sizes due to the area cuts imposed during the dwarf selection and we may have some bias against dIrrs in the sample.

The candidates in our final catalogue appear to be a robust dwarf sample, based on multiple tests that were applied throughout the paper. We first obtained distances for $\sim13$\% of the sample by cross-matching our catalogue against the SDSS database, the Updated Nearby Galaxy Catalogue, the Catalogue of Visually Classified Dwarfs, the ALFALFA HI catalogue, and by extracting helio-centric velocities from the WSRT HI datacubes, which were then used to calculate absolute magnitudes of the candidates. Of the 292 candidates with some prior distance indicator, 291 were found to have $M_g \gtrsim -18$, while the remaining galaxy is less than 0.1~mag brighter than this cutoff. In addition, the physical properties plotted on the scaling relations are all consistent with dwarf populations that have been identified in the Virgo and Fornax clusters, and the colours and estimated stellar masses also suggest that we have identified a robust dwarf sample.  

We further converted the distance indicators to velocities to test the association of the dwarf candidates with the massive galaxies in the images. By matching the dwarfs with the nearest massive ETG or LTG from the ATLAS$^{3D}$ catalogue and comparing the relative velocity between the two, we show that $\sim90$\% of this subsample form a satellite population around the massive hosts. Previous work has shown that the number of dwarf satellites likely depends on several properties of the host galaxy (e.g., mass and morphology); however, if we assume that all the dwarfs are satellites and count the number within a fixed area (in physical units) around the central ETGs in each image, we find little correlation between properties of the ETGs and the number of satellites. This does not necessarily negate the previous studies, however, as the limited coverage of the pointings, distances to the systems, and properties of the hosts may have combined to wash out any such signal. 

If the sub-sample with known distances is representative of the entire dwarf sample, this may explain the morphological breakdown of our candidates. The dwarfs were morphologically classified into two broad categories (dE and dIrr) when they were classified, with the final morphology based on the majority opinion of the classifiers; $\sim75$\% of the sample was found to be dEs. Additionally, 23.2\% were identified as nucleated. The morphology of the candidates appears to correlate with the local density, such that dIrrs are preferentially (although not exclusively) located in less dense regions and dEs in higher density environments. We further confirm the observational trend that brighter dwarfs are more likely to be nucleated than fainter dwarfs.

We would also like to highlight the fact that we have identified a population of UDGs in the MATLAS images. These galaxies appear to be an extension of the dwarf population to larger physical sizes, rather than a unique class of galaxies. This sample will be explored in future work to constrain potential formation scenarios of UDGs.

This is the first paper in a series that will explore and fully characterize the properties of low surface brightness dwarfs identified in the MATLAS images. In the future, this sample can also be used to feed machine learning software for the automatic identification of dwarfs, and will place important constraints on simulations of dwarf populations around isolated massive galaxies.

\section*{Acknowledgements}
The authors would like to thank the Next Generation Virgo cluster Survey (NGVS) and the Next Generation Fornax Survey (NGFS) teams for numerous fruitful discussions, and for allowing us access to their dwarf samples for comparison purposes. We would also like the thank the referee for their constructive feedback and help improving this manuscript. SP acknowledges support from teh New Researcher PRogram (Shinjin grant No. 2019R1C1C1009600) through the National Research Foundation of Korea. EWP acknowledges support from the National Natural Science Foundation of China under Grant No. 11573002.

Based on observations obtained with MegaPrime/MegaCam, a joint project of CFHT and CEA/IRFU, at the Canada-France-Hawaii Telescope (CFHT) which is operated by the National Research Council (NRC) of Canada, the Institut National des Science de l'Univers of the Centre National de la Recherche Scientifique (CNRS) of France, and the University of Hawaii. This work is based in part on data products produced at Terapix available at the Canadian Astronomy Data Centre as part of the Canada-France-Hawaii Telescope Legacy Survey, a collaborative project of NRC and CNRS. 

Funding for the Sloan Digital Sky Survey IV has been provided by the Alfred P. Sloan Foundation, the U.S. Department of Energy Office of Science, and the Participating Institutions. SDSS-IV acknowledges
support and resources from the Center for High-Performance Computing at
the University of Utah. The SDSS web site is www.sdss.org.

SDSS-IV is managed by the Astrophysical Research Consortium for the 
Participating Institutions of the SDSS Collaboration including the 
Brazilian Participation Group, the Carnegie Institution for Science, 
Carnegie Mellon University, the Chilean Participation Group, the French Participation Group, Harvard-Smithsonian Center for Astrophysics, 
Instituto de Astrof\'isica de Canarias, The Johns Hopkins University, 
Kavli Institute for the Physics and Mathematics of the Universe (IPMU) / 
University of Tokyo, the Korean Participation Group, Lawrence Berkeley National Laboratory, 
Leibniz Institut f\"ur Astrophysik Potsdam (AIP),  
Max-Planck-Institut f\"ur Astronomie (MPIA Heidelberg), 
Max-Planck-Institut f\"ur Astrophysik (MPA Garching), 
Max-Planck-Institut f\"ur Extraterrestrische Physik (MPE), 
National Astronomical Observatories of China, New Mexico State University, 
New York University, University of Notre Dame, 
Observat\'ario Nacional / MCTI, The Ohio State University, 
Pennsylvania State University, Shanghai Astronomical Observatory, 
United Kingdom Participation Group,
Universidad Nacional Aut\'onoma de M\'exico, University of Arizona, 
University of Colorado Boulder, University of Oxford, University of Portsmouth, 
University of Utah, University of Virginia, University of Washington, University of Wisconsin, 
Vanderbilt University, and Yale University.

This research made use of Photutils, an Astropy package for detection and photometry of astronomical sources (Bradley et al. 2019).

\bibliographystyle{mnras_mod}
\bibliography{refs}

\clearpage

\label{lastpage}

\end{document}